 \documentclass[final,1p,times]{elsarticle}

\usepackage[utf8]{inputenc}         
\usepackage[english]{babel}          

\usepackage{graphicx}               

\usepackage[small,bf]{caption2}     
\usepackage{parskip}
\usepackage{titlesec}
\usepackage{amsmath}                
\usepackage{amssymb}                

\def\R{\mathbb{R}}
\def\N{\mathbb{N}}

\newdefinition{remark}{Remark}

\journal{Journal of Fluids and Structures}
\begin{document}
\begin{frontmatter}

\date{\today}                                       

\title{
Flow noise in planar sonar applications
}

\author{
Christian Henke
}
\address{ATLAS ELEKTRONIK GmbH, \\
         Sebaldsbruecker Heerstrasse 235,\\D-28309 Bremen, Germany}
\ead{
christian.henke@atlas-elektronik.com
} 

\begin{abstract}
In this paper an investigation of flow noise in sonar applications is presented. 
Based on a careful identification of the dominant coupling effects,
the acoustic noise at the sensor position 
resulting from the
turbulent wall pressure fluctuations
is modelled with a system of 
hydrodynamic, bending and acoustic waves.
We describe for the first time an analytical solution of the problem which is 
based on a biorthogonal system which can be solved in the spectral domain without the usual simplifying assumptions.
Finally, it is demonstrated that the analytical solution describes the flow noise generation and propagation mechanisms of the considered sea trials.

\end{abstract}


\end{frontmatter}

\section{Introduction}
\begin{figure}[!htbp]
\begin{center}
\includegraphics[width=0.7\textwidth]{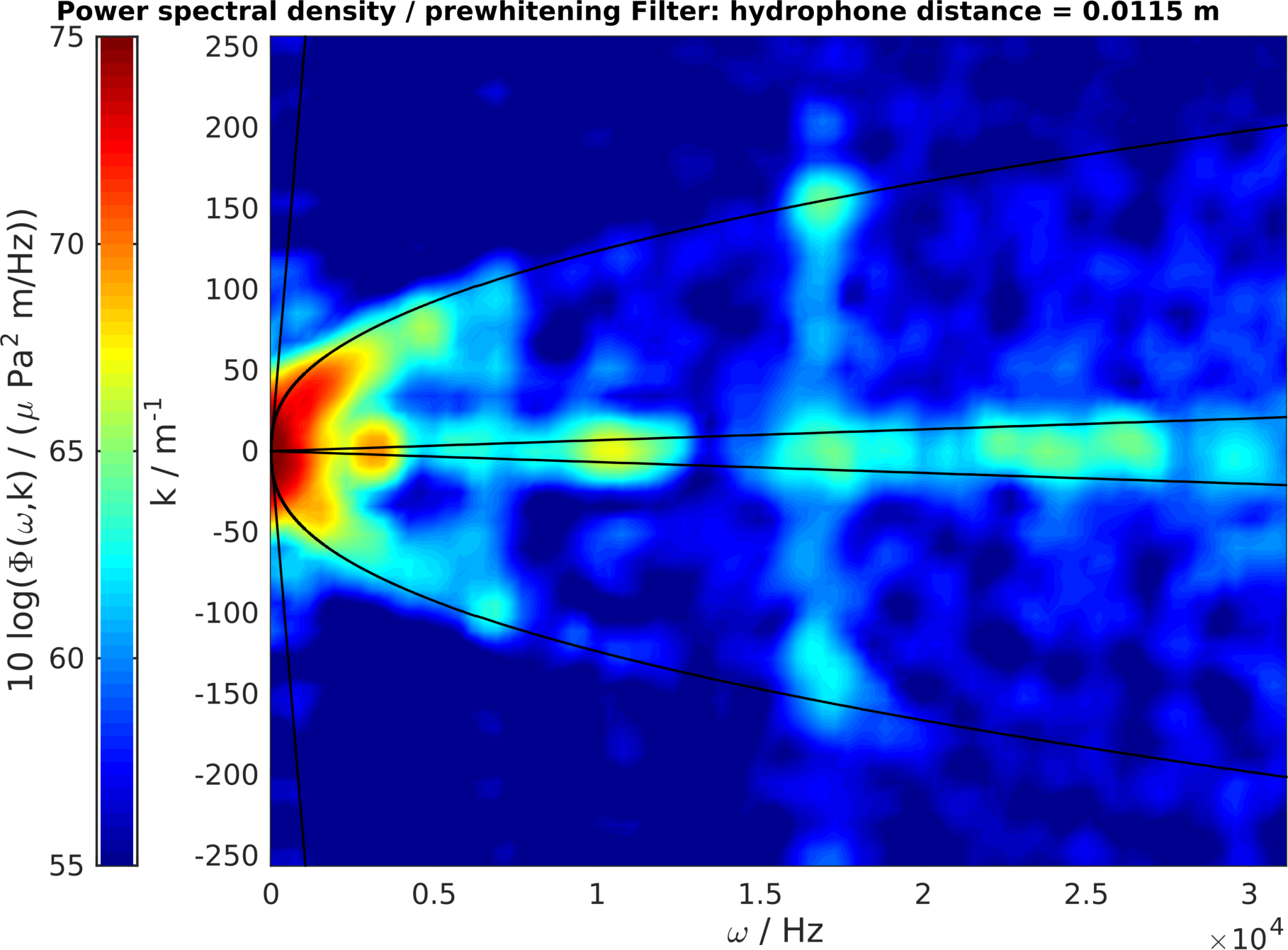}
\caption{Power spectral density of the acoustic pressure, sea trial: wavenumber-frequency plot for towing speed $u=8$kn and a hydrophone distance of $0.01$m. The straight lines show the relations $k=u^{-1} \omega$ and $k =c^{-1} \omega,$ where $k$ is the wavenumber, $\omega$ denotes the angular frequency and $c$ is the speed of sound. The curved line represents the dispersion relation which is investigated in the following sections. The dominant noise at 17000 Hz is caused by the environmental conditions. The Figure is based on a time signal of $0.2$s and is divided into segments with an overlap of $50 \%,$ which are weighted with a Flat-Top window.}
\label{fig:komega_meas}
\end{center}
\end{figure}

\begin{figure}[!htbp]
\begin{center}
\includegraphics[width=0.7\textwidth]{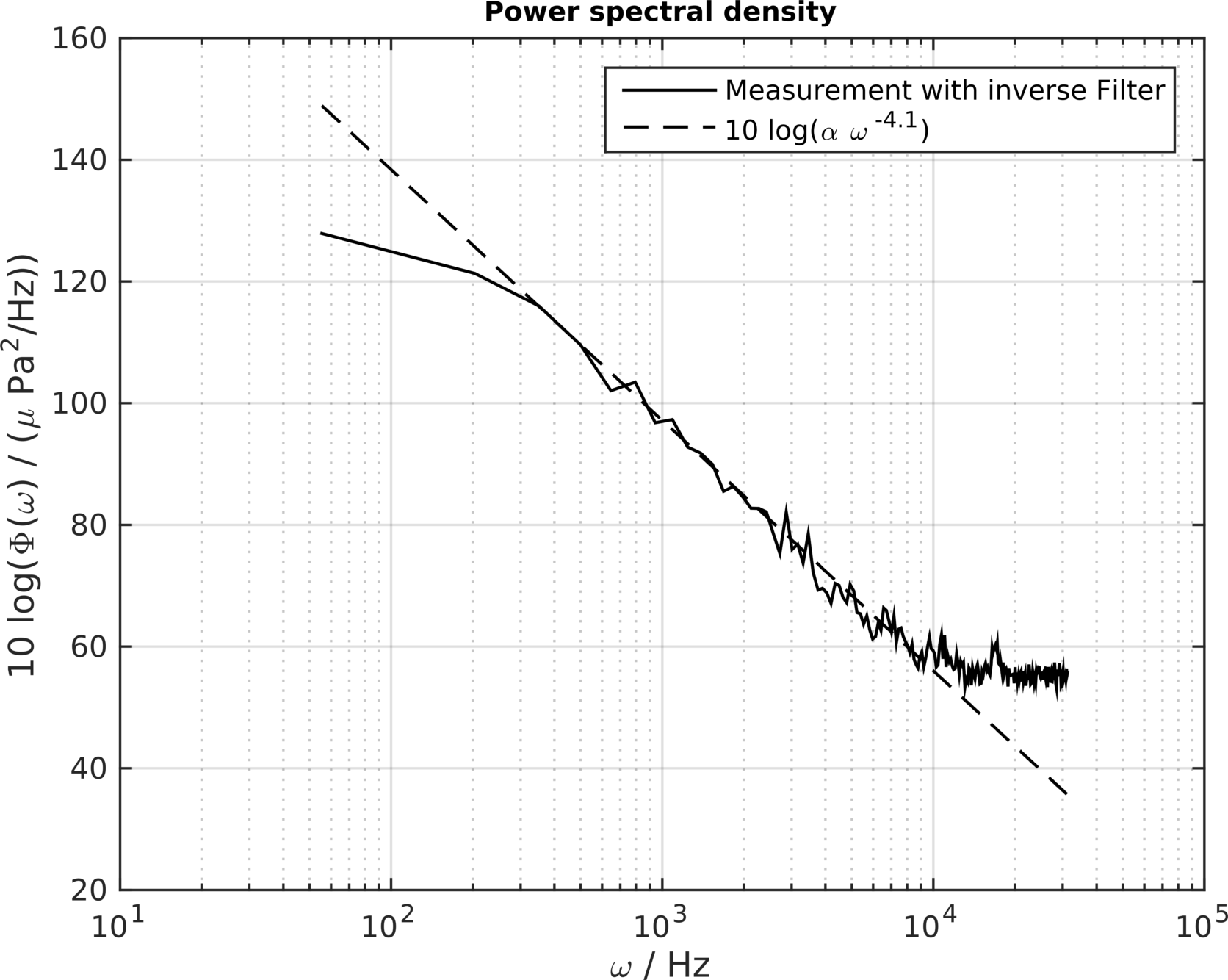}
\caption{Point pressure frequency spectrum $\Phi(\omega)=1/N \sum_{k}^N \Phi(\omega,k)$ (sea trial experiment).}
\label{fig:psd}
\end{center}
\end{figure}

Most vehicle mounted hydrophones used in passive sonar sensors are mounted inside stiff sonar windows in order to protect the sensors and to reduce the hydrodynamical drag/vibrations of the sensors.
On the one hand, when the vessel is in motion, the structure is excited by turbulent flow, and on the other hand, the 
generated turbulence causes additional noise.
Thus, the acoustic waves of these contributions propagate to the position of the hydrophone and disturb the incoming signal.
The aim of the paper is an investigation of flow noise generation and transport mechanisms with respect to different materials and geometries. The knowledge of these mechanisms could be used to maximise the signal to noise ratio. 
The mathematical modelling and a combined numerical/analytical solution of passive sonar acoustics will be validated against experiments. To consider real life underwater scenarios, the experiments involve sea trials with a test towed body, which realises a typical broadband sonar environment.
On the starboard side, the hull of the test towed body includes a linear elastic plate, which is in front of the measurement system.
 This system  consists of a line array having equally spaced hydrophones inside a water-filled measurement box, which is designed to protect the sensors from background noise. Furthermore, the box is mounted in such a way that it is decoupled from the body frame. In this way noise from the environmental conditions could be minimised. 
The flow noise generation and transport mechanisms of the sonar configuration under consideration are investigated by a wavenumber-frequency analysis. 
To reduce the dynamic range of the signal a prewhitening filter was applied to the measurement data. The filter considered the sea state, the sensor electronics and the receiving sensitivity of the hydrophones. Since the position dependent receiving sensitivity of the hydrophones is not included, position or wavenumber dependent measurement results should only be interpreted qualitatively.
Figure \ref{fig:komega_meas} and \ref{fig:psd} show the unpublished results of a sea trial experiment which has been described in \cite{Abshagen.Kueter.ea_Flowinducedinterior_2016}. In contrast to the polyurethane/fibre-reinforced plastic sandwich plate of \cite{Abshagen.Kueter.ea_Flowinducedinterior_2016}, the experimental results of this work refer to a steel plate in front of the hydrophones. 

In literature there are many investigations of sound radiated by boundary layer driven structures (see for an overview, e.g. \cite{Crighton.Dowling.ea_ModernMethods_1992,Howe_AcousticsofFluid-Structure_1998,Ciappi.DeRosa.ea_Flinovia_2015} ). 
The coupling between the boundary layer pressure fluctuation of a finite rectangular plate and the acoustic pressure fluctuation in an acoustic indefinite domain
was investigated by Strawderman \cite{Strawderman_WavevectorFrequencyAnalysis_1994} with a coupled eigenfunction expansion method. Because of the unknown inverse of the system matrix (cf. \cite[p. 5-46 - 5-47]{Strawderman_WavevectorFrequencyAnalysis_1994}), this approach cannot be used for practical applications without simplifying assumptions on the fluid loading and the crosscoupled modal terms.
Besides the above considered application, this problem is also a common task in aircraft cabin noise studies.
For aircraft applications, Graham \cite{Graham_HighFrequencyVibration_1995,Graham_BoundaryLayerInduced1_1996,Graham_BoundaryLayerInduced2_1996} proposed also a model 
for the acoustic response of a finite rectangular plate in an acoustic indefinite domain. 
These investigations were based on the empirical Efimtsov model \cite{Efimtsov_Characteristics_1982} for the forcing pressure and derive 
the wavenumber-frequency spectrum $\Phi(\omega,k)$ which is the Fourier transformation of the space correlation function of the wall pressure. 
However, in the averaging process of the Efimtsov model and comparable approaches like Corcos or Chase (see \cite[Chapter 3]{Howe_AcousticsofFluid-Structure_1998} for an overview),
much of the relevant informations (e.g. phase relationships) are lost. 

Since the above mentioned simplifications were also used by Graham, 
the approximation of the acoustic response was improved in 
\cite{Maury.Gardonio.ea_WavenumberApproachModelling1_2002,Maury.Gardonio.ea_WavenumberApproachModelling2_2002} 
by using a different set of eigenmodes.
To overcome the simplifying assumptions, Mazzoni solves in \cite{Mazzoni.Kristiansen_FiniteDifferenceMethodAcouticRadiation_1998,Mazzoni_EfficientApproximationVibroAcoutic_2003} the underlying partial differential system by a finite difference discretisation.
In \cite[Chapter 17]{Crighton.Dowling.ea_ModernMethods_1992} an asymptotic analysis is performed, which causes the author to comment that the above problem is not already well understood:
\begin{quote}
Our aim here is to show that fluid loading cannot be characterised once and for all as either \glqq light\grqq \, or \glqq heavy\grqq \, even at a fixed frequency, in a given configuration and for fixed values of the material constants for the fluid and the structure.
\end{quote}

The novelty of this work is to present the first exact analytical solution for the coupled problem of boundary layer driven structures that automatically takes into account the entirely fluid loading. 
This can be achieved by the use of a biorthogonal base system (see for example \cite{Lebedev_IntroductionFunctionalAnalysis_1997}) in space-time which includes the pressure eigenfunctions and the complex Fourier polynomials. 
Hence, the system matrix is block diagonal and can be easily inverted. 
Notice that the biorthogonality property is only valid on the interval $(-l/2,l/2),$ where $l$ is a plate dimension. 
The investigations of the above references are based on the interval $(0,l),$ which prevents the application of the biorthogonality argument. Therefore, the system matrix could not be inverted before. 

Moreover, the present work needs the complex-valued forcing pressure as input and preserves all phase informations between the coupling of  the hydrodynamic pressure, plate vibrations and acoustic waves. 
Choosing the wall pressure wavenumber-frequency spectrum from empirical methods like Corcos, Chase or Efimtsov as an input,
this would be not possible.

The outline of the paper is as follows. 
In section 2 the modelling approach with the underlying assumptions is described. The mentioned coupling effects are considered by a set of partial differential equations and boundary conditions.
Then, the analytical solution of the problem is given with the help of a biorthogonal expansion method in section 3 and is expanded with a damping mechanism in section 4. 
Moreover, the sensitivity of the main solution parameters are demonstrated by examples.
Finally, the general conclusions are discussed in section 5.

\section{Modelling approach}

It has been known for a long time in underwater acoustics that the volume based turbulent flow noise is negligible compared with the sound radiated by plate vibrations \cite[(9-47), (9-53)]{Blake_MechanicsofFlow-Induced_1986}.
Moreover, in order to describe the interaction of the flexible plate with the 
turbulent flow outside and the acoustic waves inside the measurement box, it is assumed that there
is no feedback of the  
plate vibrations on the turbulent flow
(single side coupling). 
Namely, the hydrodynamic pressure is added as a source term in the two side coupling of bending and acoustic waves.
The validity of this hybrid approach is investigated by the evaluation of the acoustic pressure in the wavenumber-frequency domain. This way, it is checked if the source term interpretation of the hydrodynamic 
flow leads to valid transport mechanisms of the coupled waves. 

In this approach, we have to start with a broadband hydrodynamical field in a computational domain over the sonar window.
To meet this demand, the incompressible Navier-Stokes equations are solved by a Large Eddy Simulation (LES), where the larger 
scales are resolved and the effects on the smaller scales are considered by a subgrid-scale model.
Moreover, the LES simulation requires turbulent inflow data, which is obtained by an auxiliary simulation with obstacles mounted
at the wall (cf. \cite[Section 5.2.2.]{Wagner.Huettl.ea_Large-EddySimulationAcoustics_2007}).
The aim of the LES simulation is not necessarily a detailed resolution of the turbulent boundary layer, but a representation of the wall pressure fluctuation that captures the necessary broadband effects for the coupled vibro-acoustic waves (cf. Table \ref{tab:LES}).

\begin{figure}[!htbp]
\begin{center}
\includegraphics[width=0.515\textwidth]{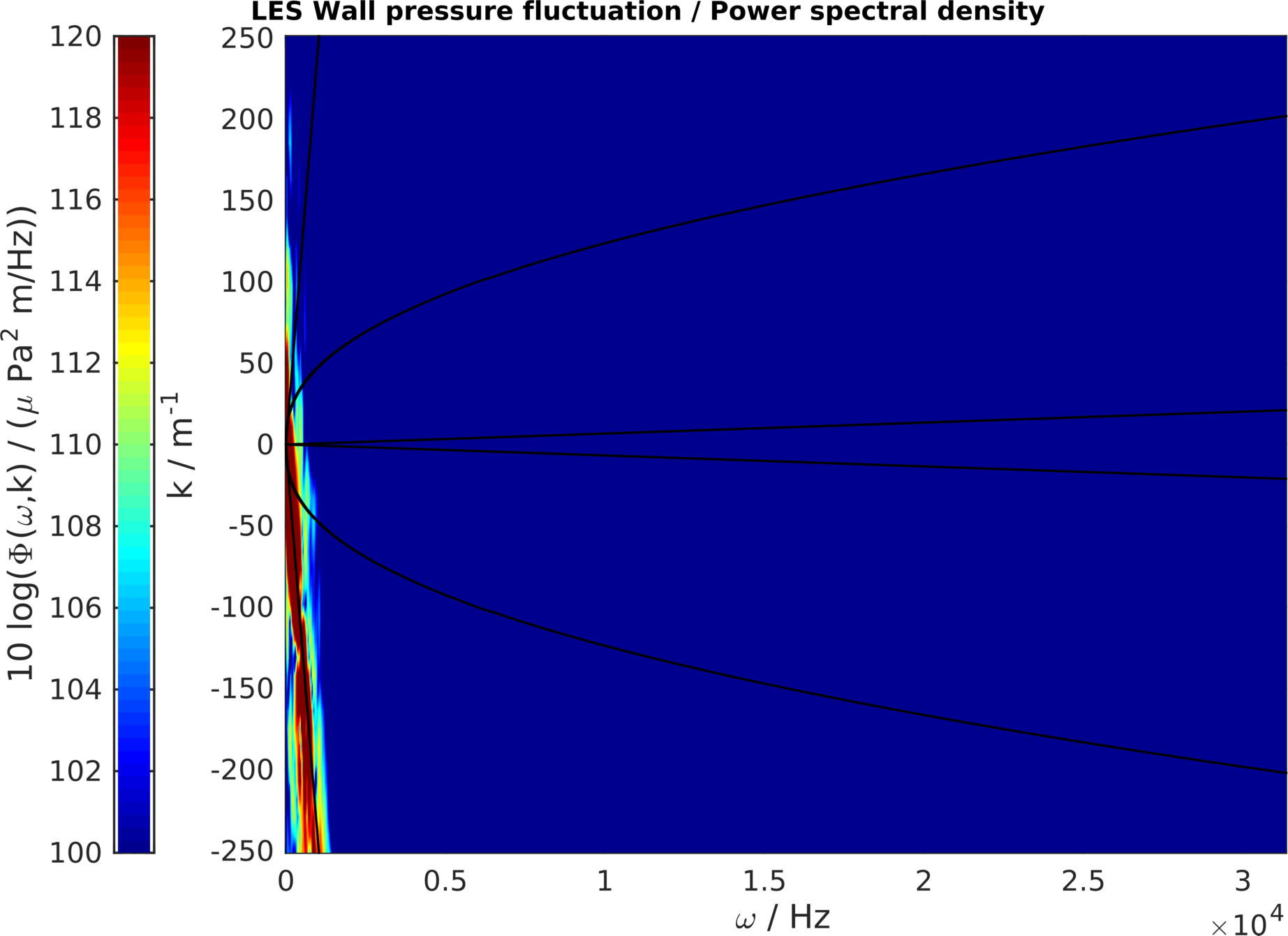}
\hfill
\includegraphics[width=0.475\textwidth]{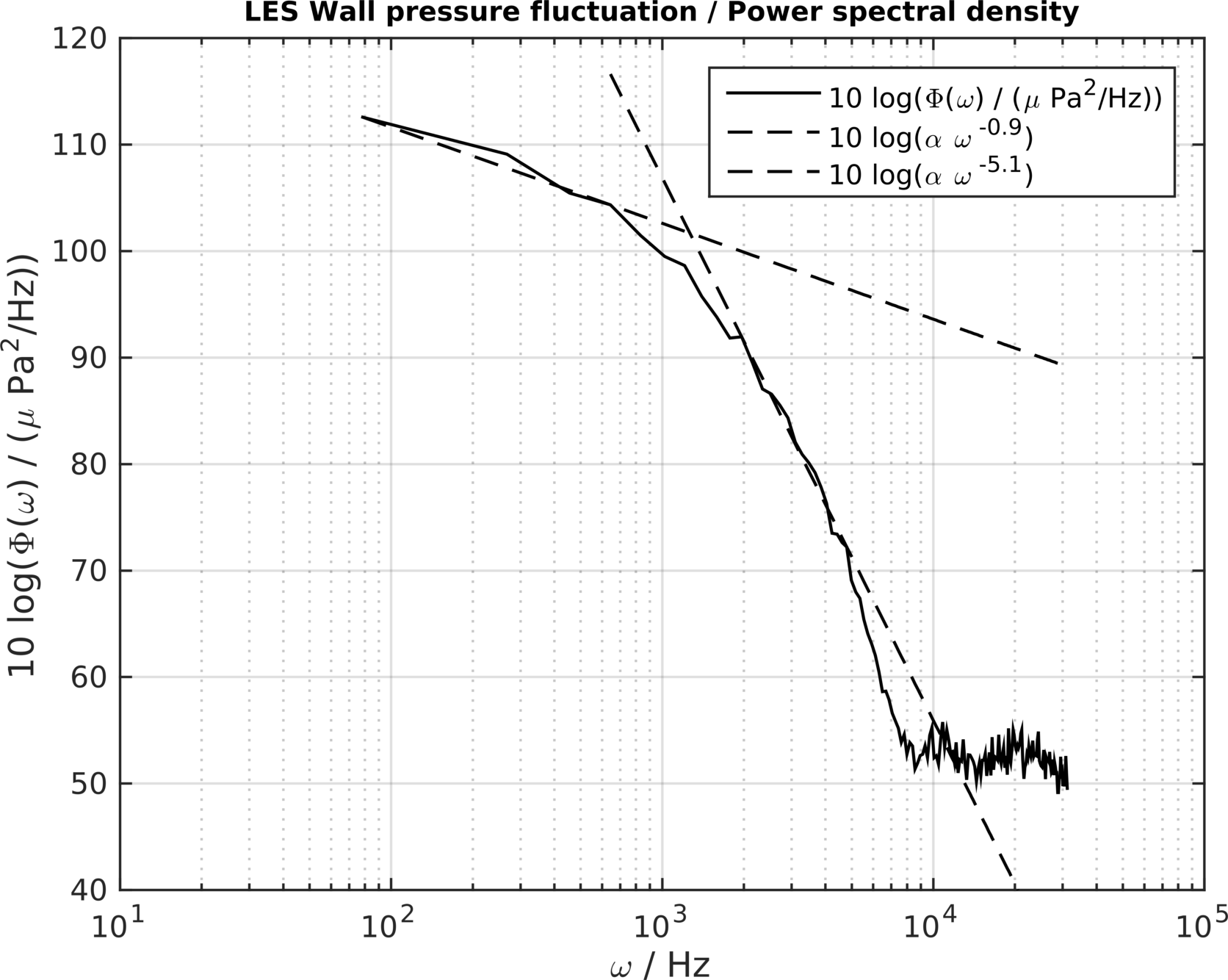}
\caption{Turbulent wall pressure spectrum of the LES simulation. Left: The convective ridge $(k=u^{-1} \omega)$ is the dominant wall pressure fluctuation, right: the LES point pressure frequency spectrum shows the correct decay behaviour of $\sim \omega^{-1}$ and $\sim \omega^{-5}$ (cf. \cite[(8-45), (8-46)]{Blake_MechanicsofFlow-Induced_1986}). The spectrum indicates that the largest frequency with a good wall pressure representation is approximately $\omega \sim 8000$Hz. }
\label{fig:hydrodynamic_psd}
\end{center}
\end{figure}

\begin{table}[h]
\begin{center}
\begin{tabular}{l|r}
LES setup & \\
\hline
Streamwise domain size, $x \,/m$ & 0.7 \\
Wall-normal domain size, $y \,/m$ & 0.2 \\
Spanwise domain size, $z \,/m$ & 0.39 \\
\hline
Kinematic viscosity $\nu \,/ 10^{-6} m^2/s$ & 1.6 \\
Towing speed $u\,/kn$ & 8 \\ 
Time step $\Delta t\, /10^{-5} s $ & 2.5 \\
Friction velocity $u_\tau\, / m/s$ &0.155 \\  
Length scale, $l^+=\nu/u_\tau\, / 10^{-6}m$ & 10.3 \\
\hline
Grid resolution & \\
$\Delta x^+_{\text{min}} = \Delta x_{\text{min}}/l^+$ & 48.44 \\
$\Delta y^+_{\text{min}} = \Delta y_{\text{min}}/l^+$ &  5.19 \\
$\Delta z^+_{\text{min}} = \Delta z_{\text{min}}/l^+$ & 48.44 \\
$\Delta x^+_{\text{max}} = \Delta x_{\text{max}}/l^+$ & 193.75 \\
$\Delta y^+_{\text{max}} = \Delta y_{\text{max}}/l^+$ & 193.75 \\
$\Delta z^+_{\text{max}} = \Delta z_{\text{max}}/l^+$ & 193.75 \\
\hline
Average growth rate in $y$ & 1.18 \\
Number of grid points $/ 10^6$ & 22\\
\end{tabular}
\caption{LES setup of the boundary layer grid}
\label{tab:LES}
\end{center}
\end{table}

Next, the space-time wall pressure fluctuations are transformed into the wavenumber-frequency domain. To achieve a good amplitude representation of the data the Flat-Top window function is chosen.

Finally, the coupling of the plate vibrations with the acoustic field has to be solved. 
Here, it is assumed that the vibrations of a plate with thickness $h$ and displacements $w$ satisfy the following condition
\begin{equation}
w \ll h \ll  \text{wavelength of the bending waves}.
\label{eq:assumption1}
\end{equation}  
Then the displacement $w$ takes place only in the perpendicular direction.
Therefore, the acoustic interaction appears only in the same direction. Consequently, the tangential domain boundaries and the associated acoustic boundary conditions in the measurement box are of minor importance and could be considered as a sound hard wall on a rectangular wall. To represent the material properties of the measurement box, the wall on the back side is equipped with an impedance boundary condition.

Next, the mathematical problem and a connected numerical simulation of the coupled acoustic and bending wave problem are presented.

We start by introducing the modelling domain, which was already mentioned in a loose way in the introduction. 
Let the measurement box be a rectangular domain $\Omega=(-l_1/2,l_1/2) \times (-l_2/2,l_2/2) \times (-l_3/2,l_3/2)$ with the side lengths $l_1,l_2,l_3.$  
The plate with uniform thickness $h$ is located at $\Gamma=\left\{ x \in \Omega: x_3=0 \right\}.$
Moreover, the hydrophones are mounted on the line $H=\left\{ x \in \Omega: 
x_2=0, x_3=l \right\},$ 
where 
$l<0$ is a given parameter. The time interval under consideration is $[0,T], T>0.$

As usual in linear, isotropic elasticity, the material parameters of the plate are described by the density $\rho_p,$ the Young modulus $E$ and the Poisson ratio $\nu.$

\begin{figure}[tbph]
\begin{center}
\includegraphics[width=0.8\textwidth]{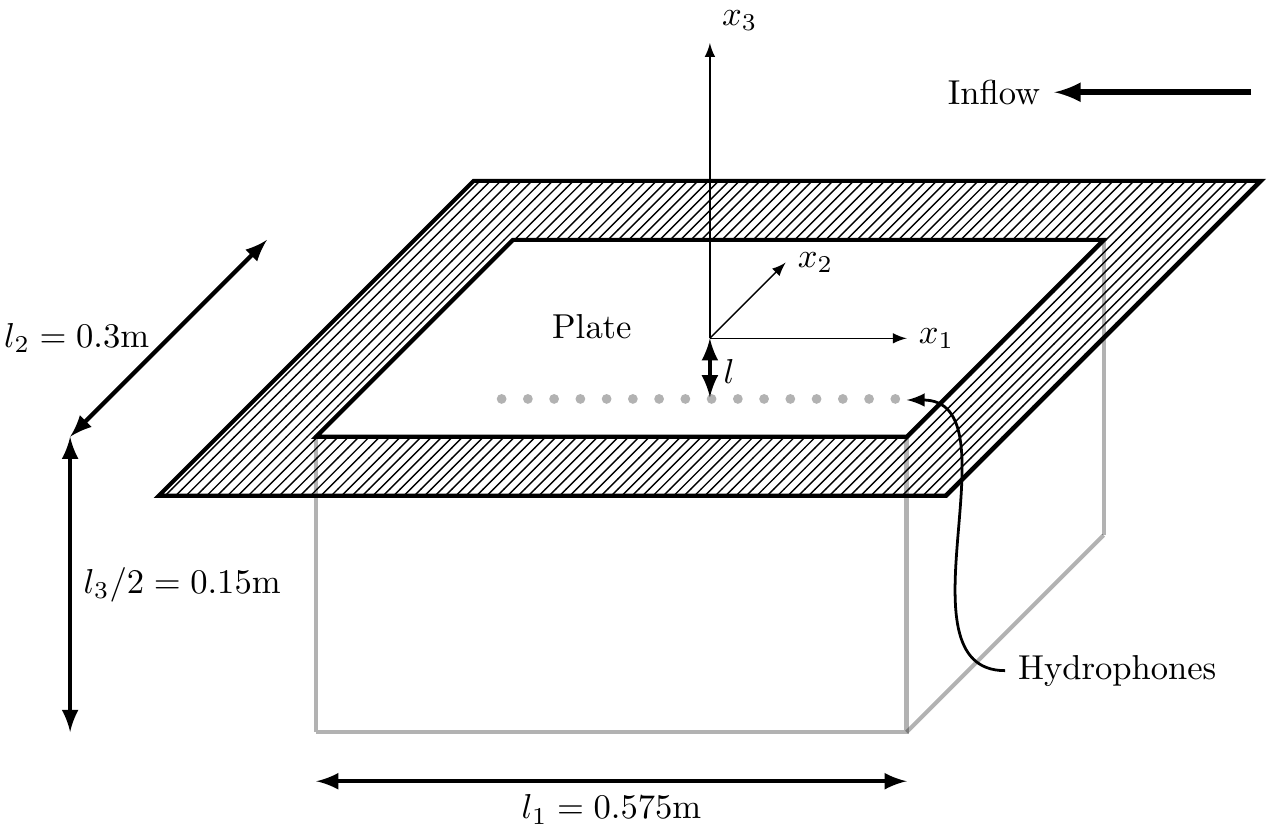}
\caption{A schematic view of the modelling domain.}
\end{center}
\end{figure}

Taking the assumption $(\ref{eq:assumption1}),$ 
the complexity of linearised elasticity can be reduced by an application of the bending wave equation
\begin{equation}
m \partial_t^2 w + B \Delta \Delta w = -(p^+ -p^-),\quad \text{on } [0,T] \times \Gamma,
\label{eq:bending_wave}
\end{equation} 
where $\Delta$ denotes the Laplace operator. Furthermore, $m=\rho_p h$ is the mass per unit area of the plate,
\begin{equation}
B=\frac{E h^3}{12(1-\nu^2)},
\label{}
\end{equation}
is the bending stiffness and
$p^+,p^-$ are the pressures above and below the plate \cite[p. 17]{Howe_AcousticsofFluid-Structure_1998}.
Namely, $p^+$ corresponds to the hydrodynamic source pressure $p_{\text{hyd}}$ plus an outgoing acoustic pressure $p_{\text{out}}$ and $p^-$ to the acoustic pressure inside the measurement box,
which consists of an ingoing wave $p_{\text{in}}$ and its reflection from the rear wall $p_{\text{refl}}.$
As usual, the ingoing and outgoing pressures are connected by 
\begin{equation}
p_{\text{out}}(t,x_1,x_2,0)=-p_{\text{in}}(t,x_1,x_2,0).
\label{eq:wave_in_out}
\end{equation}
Moreover, $p^-$ satisfies the usual wave equation
\begin{equation}
- \frac{1}{c^2} \partial_t^2 p^- + \Delta p^-=0, \quad \text{on } [0,T] \times 
 \{x \in \Omega: x_3<0\},
\label{eq:wave_eq}
\end{equation}
which interacts additionally on $[0,T] \times \Gamma$ by
\begin{equation}
\partial_t^2 w =-\frac{1}{\rho_w} \partial_{x_3}p^-.
\label{eq:coupling}
\end{equation}
Here, $c$ and $\rho_w$ denotes the speed of sound and the density of water.
It remains to specify the boundary conditions for the plate and the measurement box.
The wall displacement satisfies the plate boundary conditions
\begin{equation}
\begin{split}
w\big(t,-l_1/2,x_2\big)&=w\big(t,l_1/2,x_2\big)=
w\big(t,x_1,-l_2/2\big)=w\big(t,x_1,l_2/2\big)=
0, \\
\partial_{x_1}^2 w\big(t,-l_1/2,x_2\big)&=\partial_{x_1}^2 w\big(t,l_1/2,x_2\big)=0 \,\text{(simply supported)},\\ 
\partial_{x_2}^2 w\big(t,x_1,-l_2/2\big)&=\partial_{x_2}^2 w\big(t,x_1,l_2/2\big)=0 \,\text{(simply supported)},
\label{eq:biege_rb_ss}
\end{split}
\end{equation}
and the pressure conditions for the side and rear walls
\begin{equation}
\begin{split}
\partial_{x_1} p\big(t,-l_1/2,x_2,x_3\big)&=\partial_{x_1} p\big(t,l_1/2,x_2,x_3\big)=0 \,\text{(sound hard wall)},\\ 
\partial_{x_2} p\big(t,x_1,-l_2/2,x_3\big)&=\partial_{x_2} p\big(t,x_1,l_2/2,x_3\big)=0 \,\text{(sound hard wall)},\\
\rho_w^{-1}\partial_{x_3} p\big(t,x_1,x_2,-l_3/2\big)&=-z^{-1}\partial_t p\big(t,x_1,x_2,-l_3/2\big)=0 \,\text{(impedance)},
\label{eq:side_wall_rb}
\end{split}
\end{equation}
where $z$ denotes the acoustic impedance.

\section{Analytic solution of the acoustic/bending wave problem}
In this section, the analytical solution of the two side coupled problem is demonstrated.
Thanks to the rectangular domain, 
the displacement and the acoustic pressure can be written as a linear combination of product basis functions
\begin{equation}
\begin{split}
w(t,x_1,x_2)&=\R\left\{\sum_{n_0,n_1,n_2 \in \mathbb{Z}} w_n w_{n_0}(t)w_{n_1}(x_1)w_{n_2}(x_2)\right\} \quad \text{on } [0,T] \times \Gamma,\\
p^-(t,x_1,x_2,x_3)&=\R\left\{\sum_{n_0,n_1,n_2,n_3 \in \mathbb{Z}} p_n p_{n_0}(t)p_{n_1}(x_1)p_{n_2}(x_2)p_{n_3}(x_3)\right\} \quad \text{on } [0,T] \times \Omega.\\
\end{split}
\label{eq:tensor_prod}
\end{equation}
To prepare the frequency/wavenumber analysis of the hydrophone signals, we
assume that $w_{n_i},p_{n_i},$ $i=0,\dots,d,$ $n_0=-f,$ are periodic functions
and introduce the notations
\begin{equation*}
\langle v,w \rangle =\int_{-\frac{T}{2}}^{\frac{T}{2}} \int_{-\frac{l_1}{2}}^{\frac{l_1}{2}} 
 \int_{-\frac{l_2}{2}}^{\frac{l_2}{2}}v(t,x_1,x_2) \overline{w(t,x_1,x_2)} \, dt dx_1 dx_2, \quad \|v\|=\sqrt{\langle v,v \rangle},
\end{equation*}
where $v,w:[0,T] \times \Gamma  \to \R$ are also periodic functions.
Now, the Fourier series representation can be written as follows
\begin{equation}
\begin{split}
v(t,x_1,x_2)&=\R \left\{\sum_{n \in \mathbb{Z}^3} \frac{\langle v,\Psi_n \rangle }{\|\Psi_n\|^2} \Psi_n \right\}, \\
\Psi_n(t,x_1,x_2)&=\exp\left(i2 \pi \left(-\frac{f t}{T}+\frac{n_1 x_1}{l_1}+\frac{n_2 x_2}{l_2}\right)\right).
\end{split}
\label{eq:fourier_series}
\end{equation}
Then, using integration by parts, it is straightforward to verify for $v:[0,T] \times \Gamma$ the identities
\begin{equation}
\begin{split}
\langle \partial_t v, \Psi_n \rangle &= i k_0 \langle v,\Psi_n \rangle,\, k_0=-\frac{2 \pi f}{T}= -\omega,\\
\langle \partial_{x_j}^\alpha v, \Psi_n \rangle &= (i k_j)^\alpha \langle v,\Psi_n \rangle,\,k_j=\frac{2 \pi n_j}{l_j},\,j=1,2,\, \alpha \in \N,\\
\langle \Delta v, \Psi_n \rangle &= -k^2  \langle v,\Psi_n \rangle,\, k^2=k_1^2+k_2^2.\\
\end{split}
\label{eq:fourier_id}
\end{equation}

\subsection{Eigenfunctions of the acoustic and bending waves}
Using the product basis, one can construct eigenfunctions which satisfy the correct boundary conditions, by one-dimensional arguments.

Substituting an approach of the form
\begin{equation}
\begin{split}
w_{n_0}(t)&=\exp\big(-i \omega t\big),\\
w_{n_j}(x_j)&=\sin\big(\tilde{k}_j x_j\big),\, \tilde{k}_j=\frac{n_j \pi}{l_j},\,j=1,2,\\
\end{split}
\end{equation}
in $(\ref{eq:tensor_prod}),$ 
the displacement $w$ solves the homogeneous form of $(\ref{eq:bending_wave})$ if 
\begin{equation}
\omega=\pm \left( \frac{B}{m} \right)^{1/2} \tilde{k}^2,\, \tilde{k}^2=\tilde{k}_1^2 + \tilde{k}_2^2,
\label{eq:disp_bending}
\end{equation} 
is fulfilled.
Moreover, we find the acoustic eigenfunctions by separation of variables. Namely, $(\ref{eq:wave_eq})$ is satisfied by $(\ref{eq:tensor_prod})$ if
\begin{equation}
\begin{split}
p_{n_0}(t)&=\exp(-i \omega t), \\
p_{n_j}''(x_j)+k_{n_j}^2 p_{n_j}(x_j)&=0, \, j=1,2,3, \\
\frac{\omega^2}{c^2}&=k_1^2+k_2^2+k_3^2,
\end{split}
\label{eq:wave_eq_conseq}
\end{equation}
is claimed.
The boundary conditions $(\ref{eq:side_wall_rb})$ are fulfilled if 
\begin{equation}
\begin{split}
p_{n_j}'(-l_j/2)&=p_{n_j}'(l_j/2)=0,\, j=1,2,\\
\frac{1}{\rho_w}p_{n_3}'(-l_3/2)&=\frac{i \omega}{z}p_{n_3}(-l_3/2).
\end{split}
\label{eq:side_wall_rb2}
\end{equation}
is implemented.
Therefore, we obtain
\begin{equation}
p_{n_j}(x_j)=\cos(k_j x_j),\, j=1,2,
\label{eq:sound_hardwall2}
\end{equation} 
and with a short calculation based on a sum of an incident and a reflected wave, the impedance boundary condition is satisfied by
\begin{equation}
p_{n_3}(x_3)=\alpha_n \left[ 
\exp(-i k_3 x_3) 
+\frac{k_3 z +\omega \rho_w}{k_3 z -\omega \rho_w}
  \exp(i k_3 l_3) \exp(i k_3 x_3)
\right], \, \alpha_n  \in \mathbb{C}.
\label{eq:impedance_bc2}
\end{equation}  
\subsection{Solution of the coupled problem}
Inserting the eigenfunctions into $(\ref{eq:tensor_prod})$ and using $(\ref{eq:wave_eq_conseq}),$
the summation with respect to $n_3$ reduces to one term concerning $k_3=\sqrt{(\omega^2/c^2-k_1^2-k_2^2)}.$
Thus equation $(\ref{eq:tensor_prod})$ becomes 
\begin{equation}
\begin{split}
w(t,x_1,x_2)&=\R\left\{\sum_{n_0,n_1,n_2 \in \mathbb{Z}} w_n \exp(-i \omega t) \sin(\tilde{k}_1 x_1) \sin(\tilde{k}_2 x_2)\right\},\\
p^-(t,x_1,x_2,x_3)&=\R\Bigg\{\sum_{n_0,n_1,n_2 \in \mathbb{Z}} p_n^- \exp(-i \omega t) \cos(k_1 x_1) \cos(k_2 x_2) \times \\
& \quad \times\left(
\exp(-i k_3 x_3) 
+\frac{k_3 z +\omega \rho_w}{k_3 z -\omega \rho_w}
  \exp(i k_3 l_3) \exp(i k_3 x_3)
\right)
\Bigg\}.\\
\end{split}
\label{eq:tensor_prod2}
\end{equation}
Notice that the coefficient $\alpha_n$ from $(\ref{eq:impedance_bc2})$ is accounted by $p_n^-.$
Now, in order to determine the coefficients $w_n$ and $p_n^-,$ it remains to consider the two side coupling by $(\ref{eq:bending_wave})$ and $(\ref{eq:coupling}).$
To do so, we define
\begin{equation}
\begin{split}
\Phi_n(t,x_1,x_2)&=\exp(-i \omega t) \sin(\tilde{k}_1 x_1) \sin(\tilde{k}_2 x_2),\\
\Theta_n(t,x_1,x_2)&=\exp(-i \omega t) \cos(k_1 x_1) \cos(k_2 x_2),\\
\Theta_{n_3}(x_3)&=\exp(-i k_3 x_3) 
+\frac{k_3 z +\omega \rho_w}{k_3 z -\omega \rho_w}
  \exp(i k_3 l_3) \exp(i k_3 x_3),
\end{split}
\label{eq:eigenfunctions}
\end{equation}
and write the two side coupling for every $m \in \mathbb{Z}^3$ in the following manner:

\begin{equation}
\begin{split}
m \langle \partial_t^2 w, \Psi_m \rangle + B \langle \Delta \Delta w, \Psi_m \rangle - \langle p^-,\Psi_m \rangle 
&=-\langle p^+,\Psi_m \rangle, \\
\langle \partial_t^2 w, \Psi_m \rangle
+\frac{1}{\rho_w} \langle \partial_{x_3} p^-, \Psi_m\rangle&=0.
\end{split}
\label{eq:galerkin}
\end{equation}
Using $(\ref{eq:fourier_id}),$ it follows 
\begin{equation*}
\begin{split}
\left(-m \omega_m^2  + B \|k_m\|^4 \right)\langle w, \Psi_m  \rangle - \langle p^-,\Psi_m \rangle 
&=-\langle p^+,\Psi_m \rangle, \\
- \omega_m^2\langle w, \Psi_m \rangle
+ \frac{1}{\rho_w} \sum_{n \in \mathbb{Z}^3} p_n^- \Theta'_{n_3}(0)  \langle \Theta_n, \Psi_m\rangle&=0,
\end{split}
\label{}
\end{equation*}
and furthermore by the biorthogonal identity
\begin{equation}
\begin{split}
\langle \Theta_n,\Psi_m \rangle
&=\left( \frac{T}{2 \pi} \int_{-\pi}^{\pi} \exp(-i n_0t') \exp(im_0t')\, dt'\right)
\left( \frac{l_1}{2 \pi} \int_{-\pi}^{\pi} \cos(n_1 y_1) \exp(-im_1 y_1)\, dy_1 \right) \times \\
&\times \left( \frac{l_2}{2 \pi} \int_{-\pi}^{\pi} \cos(n_2 y_2) \exp(-im_2 y_2)\, dy_2 \right) \\
&=
T l_1 l_2 
\begin{Bmatrix}
1, & m_0=n_0=0\\
1, & m_0=n_0\neq0\\
0,& m_0 \neq n_0
\end{Bmatrix}
\begin{Bmatrix}
1, & m_1=n_1=0\\
\frac{1}{2}, & m_1=n_1\neq0\\
0,& m_1 \neq n_1
\end{Bmatrix}
\begin{Bmatrix}
1, & m_2=n_2=0\\
\frac{1}{2}, & m_2=n_2\neq0\\
0,& m_2 \neq n_2
\end{Bmatrix}
\end{split}
\label{eq:biortho}
\end{equation}
and $(\ref{eq:wave_in_out})$ it follows that
\begin{equation*}
\begin{split}
\left(-m \omega_m^2  + B \|k_m\|^4 \right)\langle w, \Psi_m  \rangle - \langle p^-+p_{\text{in}},\Psi_m \rangle 
&=-\langle p_{\text{hyd}},\Psi_m \rangle, \\
- \omega_m^2\langle w, \Psi_m \rangle
+ \frac{1}{\rho_w} p_m^- \Theta'_{m_3}(0)  T l_1 l_2 
\begin{Bmatrix}
1, & m_1=0\\
\frac{1}{2}, & m_1\neq0\\
\end{Bmatrix}
\begin{Bmatrix}
1, & m_2=0\\
\frac{1}{2}, & m_2\neq0\\
\end{Bmatrix}
&=0.
\end{split}
\end{equation*}
Notice that the biorthogonal identity $(\ref{eq:biortho})$ is not valid if the inner-product is based on the interval $(0,T) \times (0,l_1) \times (0,l_2).$
Consequently, by using the biorthogonal identity $(\ref{eq:biortho})$ again, we have the representation
\begin{equation*}
p_m^-
=-\left[ \left(-m \omega_m^2  + B \|k_m\|^4 \right)\frac{\Theta'_{m_3}(0)}{\rho_w \omega_m^2} -  \Theta_{m_3}(0) -1 \right]^{-1} \frac{\langle p_{\text{hyd}},\Psi_m \rangle}{T l_1 l_2}  
\begin{Bmatrix}
1, & m_1=0\\
2, & m_1\neq0\\
\end{Bmatrix}
\begin{Bmatrix}
1, & m_2=0\\
2, & m_2\neq0\\
\end{Bmatrix}
\label{}
\end{equation*}
It should be noted, that the calculation of $w_m$ runs analogously.

Thus, applying the definition $(\ref{eq:tensor_prod2}),$  a transfer function between the hydrodynamic and acoustic Fourier coefficients can be defined
\begin{equation}
\begin{split}
\widehat{p^-}(\omega_m,k_{m_1},k_{m_2},x_3)&=\frac{\langle p^-,\Psi_m \rangle}{\|\Psi_m\|^2}
=F\left(\omega_m,k_{m_1},k_{m_2},x_3 \right)\frac{\langle p_{\text{hyd}},\Psi_m \rangle}{\|\Psi_m\|^2}\\
&=F\left(\omega_m,k_{m_1},k_{m_2},x_3 \right) \widehat{p_\text{hyd}}(\omega_m,k_{m_1},k_{m_2}),
\end{split}
\label{eq:transfer_function}
\end{equation} 
where
\begin{equation}
F\left(\omega_m,k_{m_1},k_{m_2},x_3 \right)=-\left[ \left(-m \omega_m^2  + B \|k_m\|^4 \right)\frac{\Theta'_{m_3}(0)}{\rho_w \omega_m^2} -  \Theta_{m_3}(0) -1 \right]^{-1} \Theta_{m_3}(x_3).
\label{eq:transfer_function2}
\end{equation}

\begin{remark}
Even though it has been shown that the plate vibrations caused by a turbulent boundary layer can depend strongly on the boundary conditions \cite{Hambric.Hwang.ea_VibrationsPlates_2004},
the presented approach is not restricted to plates that are simply supported on all edges. 
To be more precise, a mixture of simply supported and clamped boundary conditions expands the one dimensional eigenfunctions of $w$ to a linear combination of $sin, cos, sinh$ and $cosh$ (cf. \cite[Section 2]{Xing.Liu_NewExactSolutionsFreeVibrations_2009}). Using the fact that $w$ is completely eliminated in $(\ref{eq:transfer_function2}),$ it can be concluded that the Fourier coefficients of $p^-$ are independent of the plate boundary conditions. 
\end{remark}

To get the Fourier coefficients $\widehat{p^-}$ at $x_2=0,$ we find from $(\ref{eq:fourier_series})$ 
\begin{equation}
\widehat{p^-}(\omega_m,k_{m_1},x_2=0,x_3)
=\sum_{k_{m_2}} F\left(\omega_m,k_{m_1},k_{m_2},x_3 \right)\frac{\langle p_{\text{hyd}},\Psi_m \rangle}{\|\Psi_m\|^2}.
\label{eq:acoustic_x2_sol}
\end{equation} 

\begin{remark}
In contrast to $|\widehat{p^-}(\omega_m,k_{m_1},k_{m_2},x_3)|^2=|F\left(\omega_m,k_{m_1},k_{m_2},x_3 \right)|^2 |\widehat{p_\text{hyd}}(\omega_m,k_{m_1},k_{m_2})|^2,$ the power spectral density $|\widehat{p^-}(\omega_m,k_{m_1},x_2=0,x_3)|^2$ cannot be calculated from the hydrodynamic power spectral density. Equation $(\ref{eq:acoustic_x2_sol})$ shows that the hydrodynamic phase information has to be considered.
\end{remark}

\begin{figure}[!htbp]
\begin{center}
\includegraphics[width=0.7\textwidth]{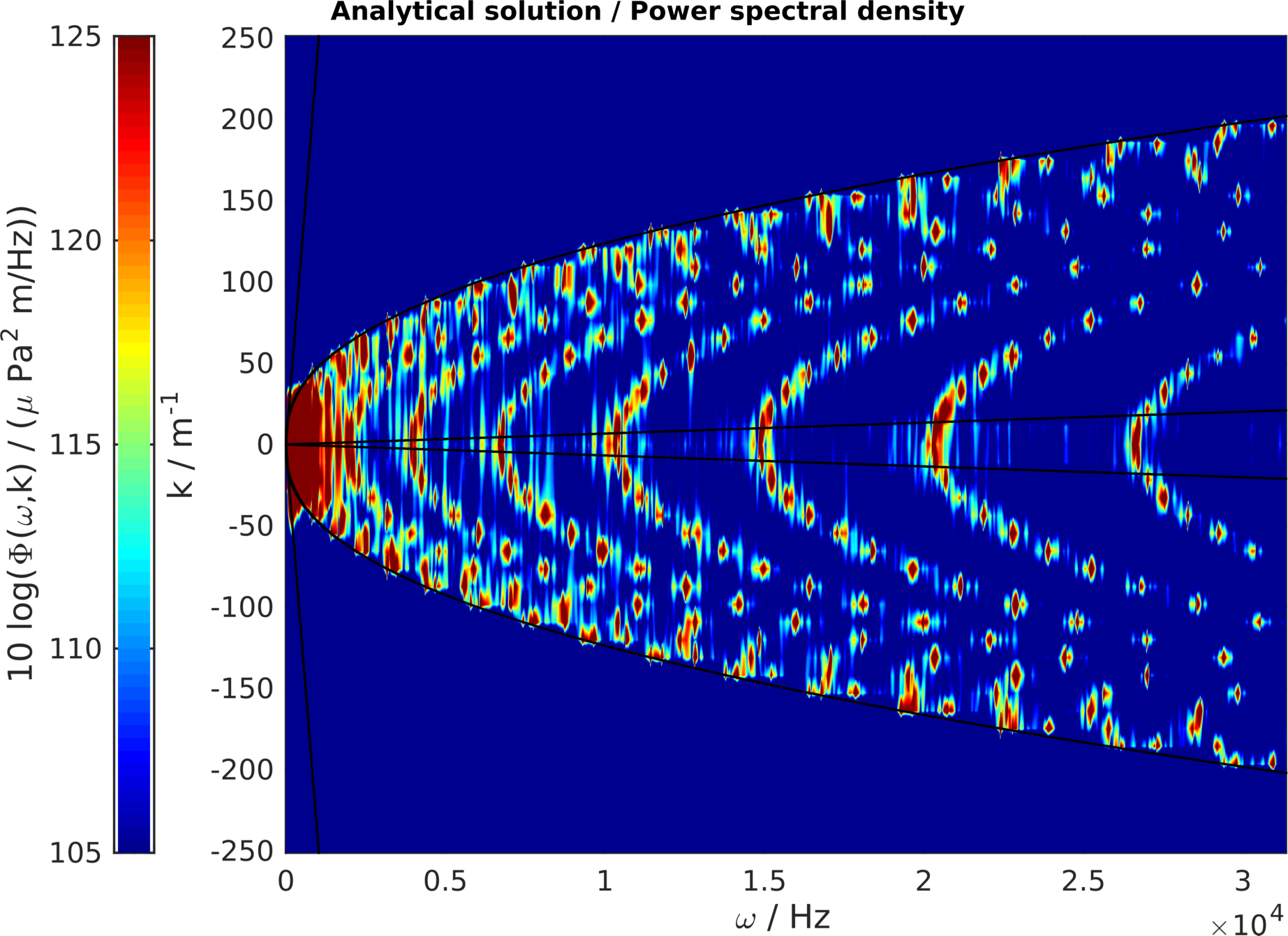}
\caption{Power spectral density $\Phi(\omega_m, k_{m_1})=\left|\widehat{p^-}(\omega_m,k_{m_1},0,0.01)\right|^2$ (analytical solution) for towing speed $u=8$kn:  wavenumber-frequency plot at $x_2=0,x_3=-0.01.$ The pressure peaks occurring in the PSD identify the eigenfrequencies of the coupled problem $(\ref{eq:bending_wave})$ and $(\ref{eq:coupling}).$}
\label{fig:komega_freq_ana}
\end{center}
\end{figure}

As a special case the transfer function without backward waves $(k_3 z +\omega \rho_w=0)$ is considered. Then it follows that 
\begin{equation*}
F\left(\omega_m,k_{m_1},k_{m_2},x_3 \right)=-\left[ \left(-m \omega_m^2  + B \|k_m\|^4 \right)\left( -\frac{i k_3}{\rho_w \omega_m^2} \right) - 2  \right]^{-1} \exp(-ik_3 x_3),
\end{equation*}
which leads for every $k_2=\frac{2 \pi n_j}{l_j}$ to the dispersion relation
\begin{equation}
-m \omega_m^2  + B \|k_m\|^4 - \frac{2 i \rho_w \omega_m^2}{k_3} =0.
\label{eq:disp_rel}
\end{equation}
Notice that the black line from Figures \ref{fig:komega_meas}, \ref{fig:komega_freq_ana} and \ref{fig:komega_freq_ana_damp} represents the dispersion relation for $k_2=0.$

However, the transfer function is only useful for theoretical observations. 
The large differences between the analytical solution and the sea trial results indicates the great importance of damping phenomena in the modelling process.
Therefore, the next section is devoted to damping effects. 

\section{Acoustic and bending waves with damping}
In this section, damping effects are added to the analytical solution of the coupled model problem $(\ref{eq:bending_wave}), (\ref{eq:wave_eq})$ and $(\ref{eq:coupling}).$ Inspired by the damped harmonic oscillator these effects are related to the undamped eigenfrequencies of the bending wave equation. 
Moreover, in the experimental configuration the plate is larger than the size of the acoustic window  $l_1 l_2$ of the measurement box. A second layer of the plate (outside the acoustic window) with steel beams and damping material defines the size of the window in the $x$- and $y$-direction, respectively. Therefore, the damping mechanism of the bending waves depends on the orthogonal directions $x,y$ and is larger in the $y$-direction. To reflect this situation we propose a generalised orthotropic damping model with damping coefficients $\eta_1$ and $\eta_2$ such that the best agreement with the experiment is ensured. 

To do that, the usual harmonic oscillator at position $x=\hat{x} \exp(-i \omega t)$ with a mass $m$ and a constant $k$ is considered
\begin{equation*}
-m \omega^2 x + k x=0.
\label{eq:harmonic_oscillator}
\end{equation*}
Now, the damped harmonic oscillator reads as follows
\begin{equation}
-m \omega^2 x - i 2 m \eta \omega \omega_0 x +k x=0,
\label{eq:damped_harmonic_oscillator}
\end{equation} 
where $\omega_0=\sqrt{k/m}$ is the eigenfrequency of the undamped oscillator and $\eta$ is a normed damping parameter. 
Considering the time-domain, the damped harmonic oscillator could be realised by the substitution $\partial_t^2 \mapsto \partial_t^2 +2 \eta \omega_0 \partial_t.$

Returning to the problem under consideration, the coupled plate vibration could be damped by applying the above substitution with the eigenfrequencies of the undamped plate
$\omega_0=\sqrt{B/m}\|k_m\|^2$ in $(\ref{eq:galerkin}).$ 
Therefore, the damped transfer function is given by
\begin{equation*}
F=
-\left[ \left(-m \omega_m^2-i2 \eta m \omega_m \omega_0  + B \|k_m\|^4 \right)\frac{\Theta'_{m_3}(0)}{\rho_w (\omega_m^2+i2 \eta \omega_m \omega_0 )} -  \Theta_{m_3}(0) -1 \right]^{-1} \Theta_{m_3}(x_3).
\label{eq:transfer_function3}
\end{equation*}
For the simplicity of notation, the arguments from the transfer function are omitted.

Now, the orthotropic damping is introduced by replacing
\begin{equation}
\eta \omega_0 
= \sqrt{\frac{B}{m}} \left( \eta k_{m_1}^2+ \eta k_{m_2}^2 \right)
\mapsto \sqrt{\frac{B}{m}} \left( \eta_1 k_{m_1}^2+ \eta_2 k_{m_2}^2 \right).
\label{eq:ana_damp}
\end{equation}

\begin{figure}[!htbp]
\begin{center}
\includegraphics[width=0.7\textwidth]{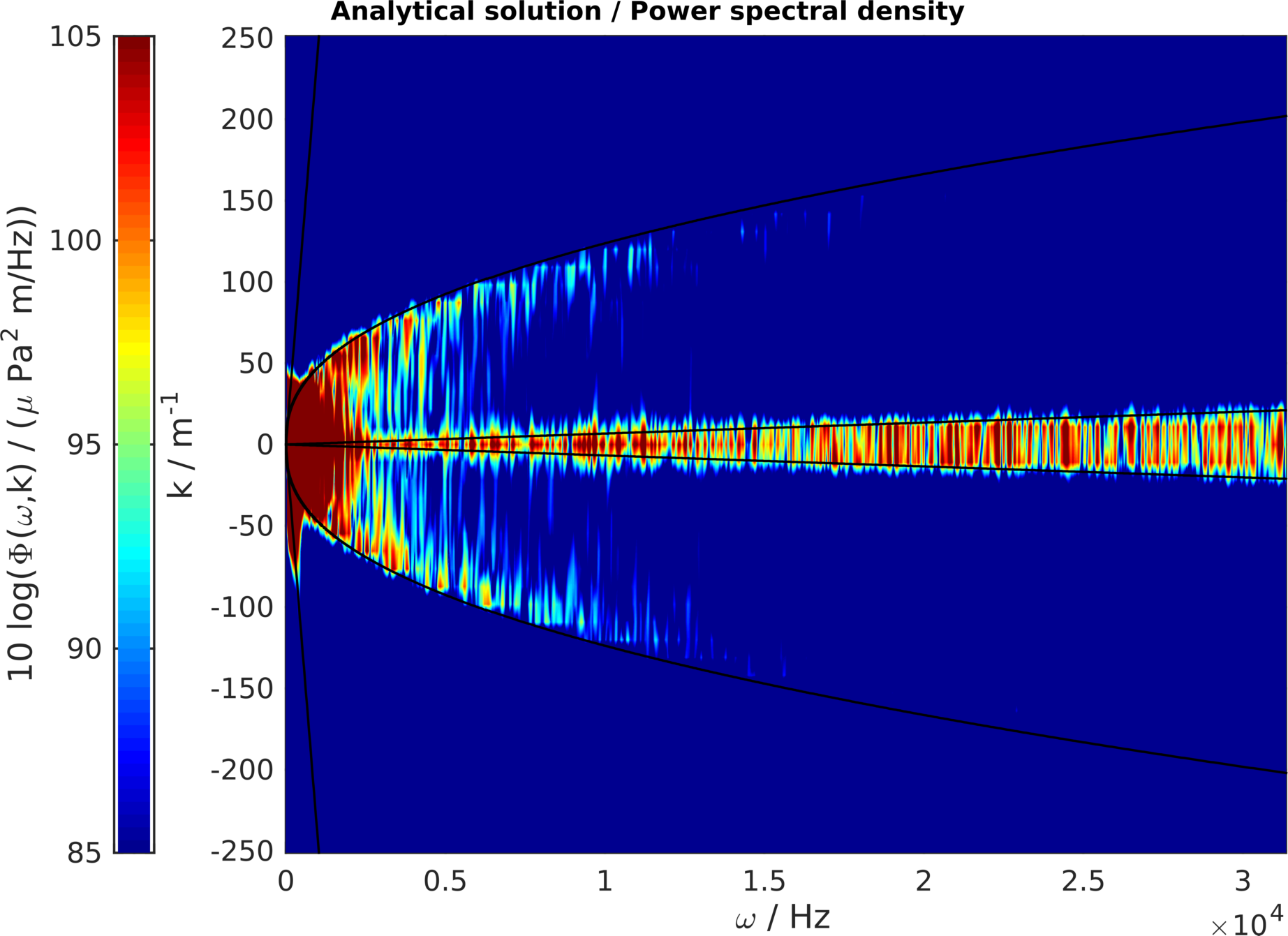}
\caption{Power spectral density $\Phi(\omega_m, k_{m_1})=\left|\widehat{p^-}(\omega_m,k_{m_1},0,0.01)\right|^2$ (analytical solution with damping $\eta_1=0.05,\,\eta_2=0.25$) for towing speed $u=8$kn: wavenumber-frequency plot at $x_2=0,x_3=-0.01.$}
\label{fig:komega_freq_ana_damp}
\end{center}
\end{figure}

\begin{figure}[!htbp]
\begin{center}
\includegraphics[width=0.7\textwidth]{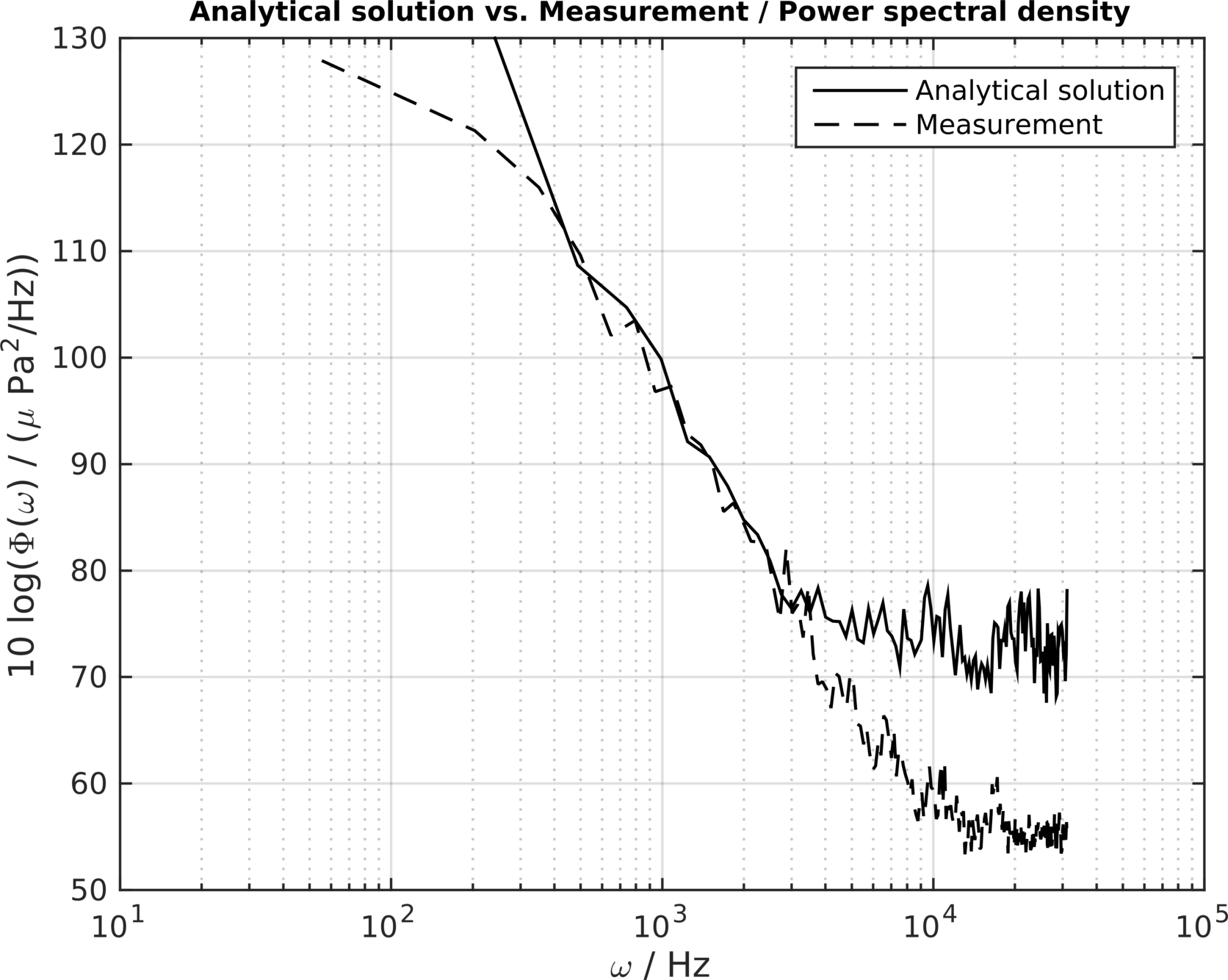}
\caption{Point pressure frequency spectrum (analytical solution with damping $\eta_1=0.05,\,\eta_2=0.25$).}
\label{fig:psd_ana_damp}
\end{center}
\end{figure}

\begin{figure}[!htbp]
\begin{center}
\includegraphics[width=0.7\textwidth]{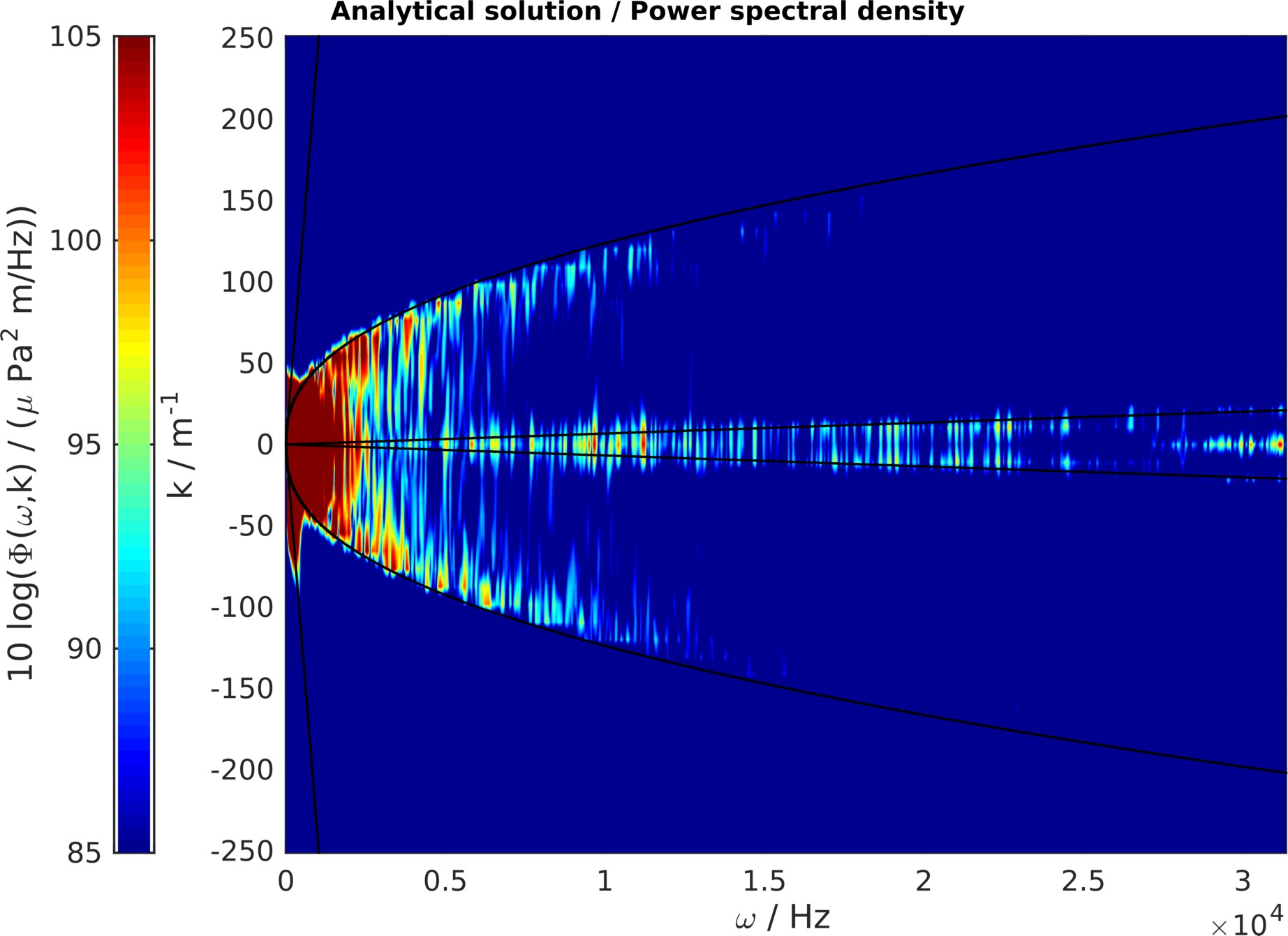}
\caption{Power spectral density $\Phi(\omega_m, k_{m_1})=\left|\widehat{p^-}(\omega_m,k_{m_1},0,0.01)\right|^2$ (analytical solution without backward waves and with damping $\eta_1=0.05,\,\eta_2=0.25$) for towing speed $u=8$kn: wavenumber-frequency plot at $x_2=0,x_3=-0.01.$}
\label{fig:komega_freq_ana_damp_wob}
\end{center}
\end{figure}

In the following the Power spectral densities of the analytical solution with orthotropic damping are presented in Figures \ref{fig:komega_freq_ana_damp}-\ref{fig:komega_freq_ana_damp_wob}.

\begin{figure}[!htbp]
\begin{center}
\includegraphics[width=0.7\textwidth]{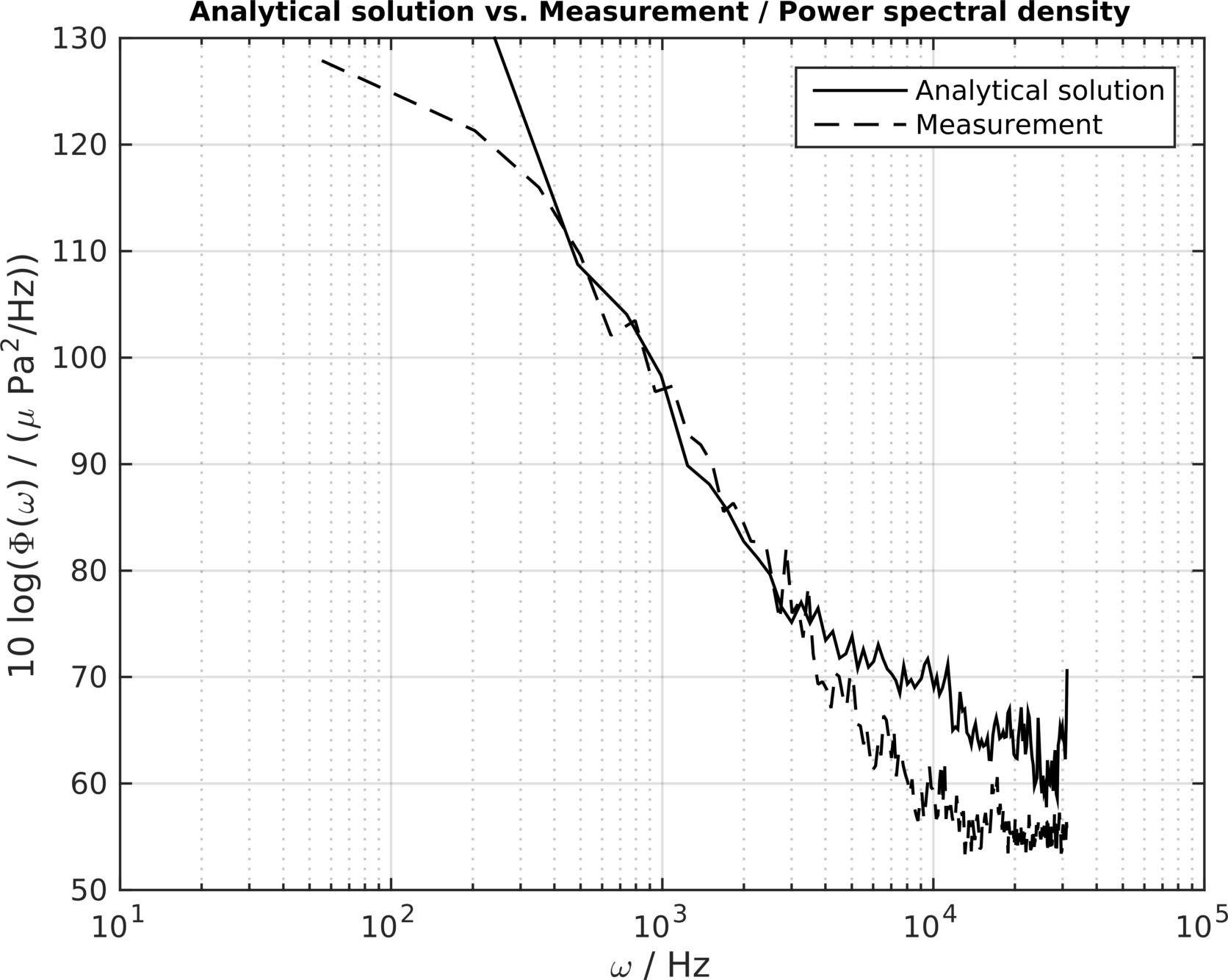}
\caption{Point pressure frequency spectrum (analytical solution without backward waves and with damping $\eta_1=0.05,\,\eta_2=0.25$).}
\label{fig:psd_ana_damp_wob}
\end{center}
\end{figure}

Figure \ref{fig:komega_freq_ana_damp_wob} indicates that there is a better agreement between the analytical solution without backward waves and the measurement. This does not seem surprising if one considers the hydrophone holder which shades the hydrophone from the backward waves.  

\begin{remark}
The subsonic $(\omega/c < \|k\|)$ and the supersonic \glqq leaky\grqq \, wave mode $(\|k\| < \omega/c)$ are automatically considered within the complex wavenumber $k_3.$ In the first case, the wave field exponentially decays with the distance to the plate, and in the second case, the wave field radiates into the fluid (acoustic domain). Here, the angle $\theta$ between the radiated wave front and the plate is given by $\sin(\theta)=\|k\|c/\omega.$ The identity $\omega/c=\|k\|$ is shown by black lines in Figures \ref{fig:komega_meas}, \ref{fig:komega_freq_ana} and \ref{fig:komega_freq_ana_damp}. 
\end{remark}

\clearpage
\subsection{Sensitivity of the solution parameters}

\begin{figure}[!htbp]

\begin{center}
\includegraphics[width=0.515\textwidth]{komega_hydrophone_pressure_ref_analytic_damp0_005_025_flattopwin_wob}
\hfill
\includegraphics[width=0.475\textwidth]{energy_spectral_density_ref_analytic_damp0_005_025_flattopwin_wob_sa}
\end{center}

\begin{center}
\includegraphics[width=0.515\textwidth]{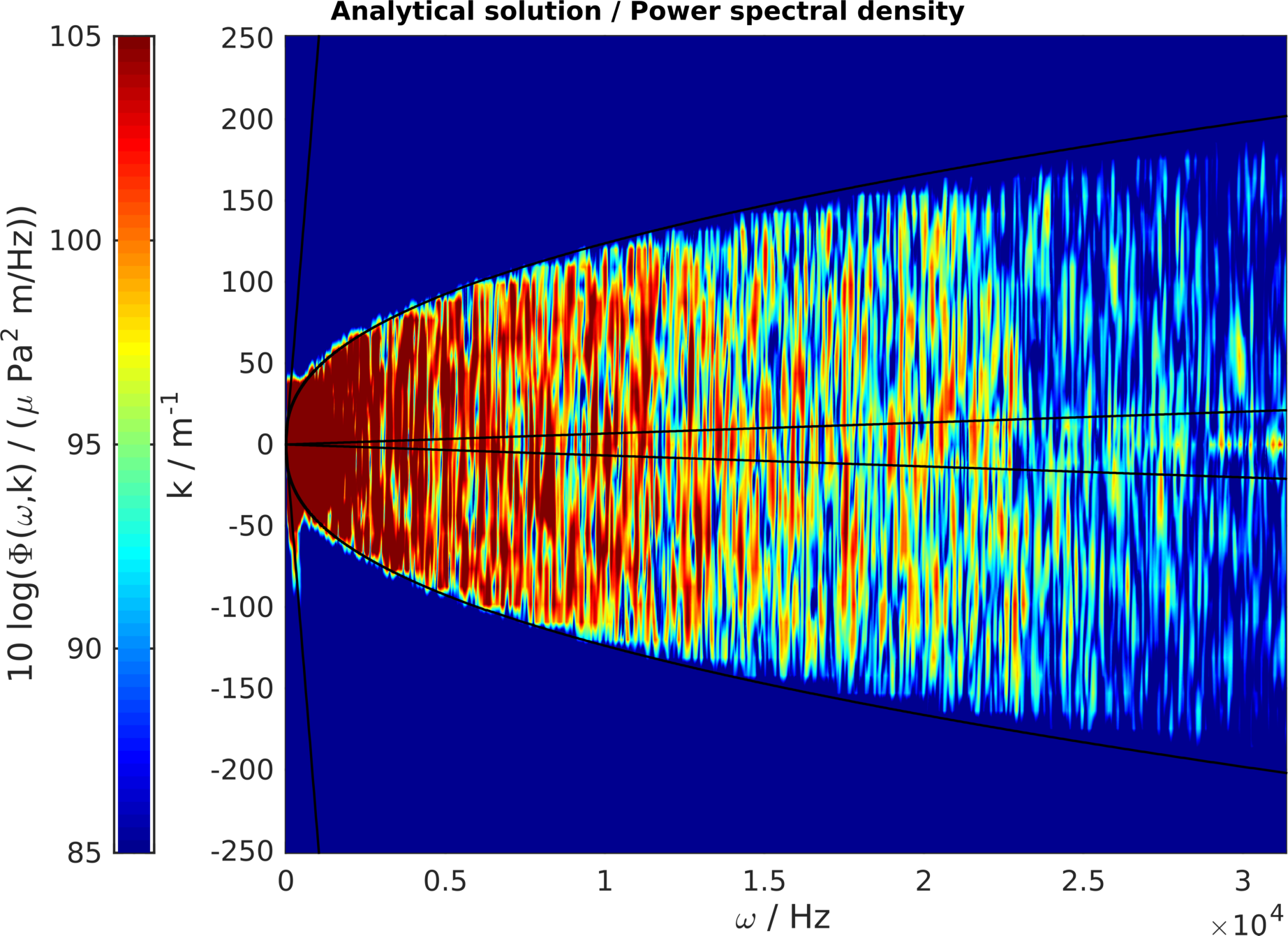}
\hfill
\includegraphics[width=0.475\textwidth]{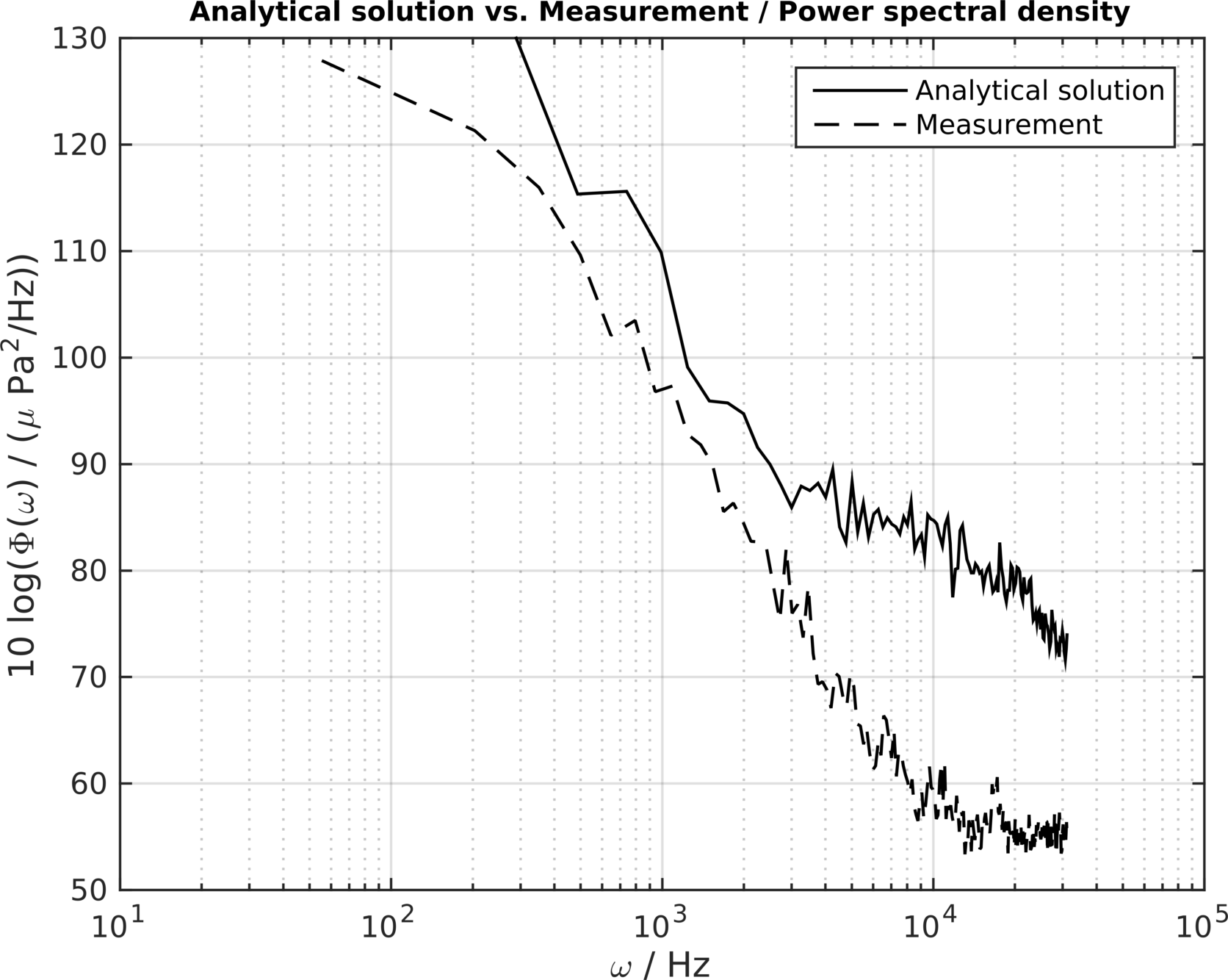}
\end{center}

\begin{center}
\includegraphics[width=0.515\textwidth]{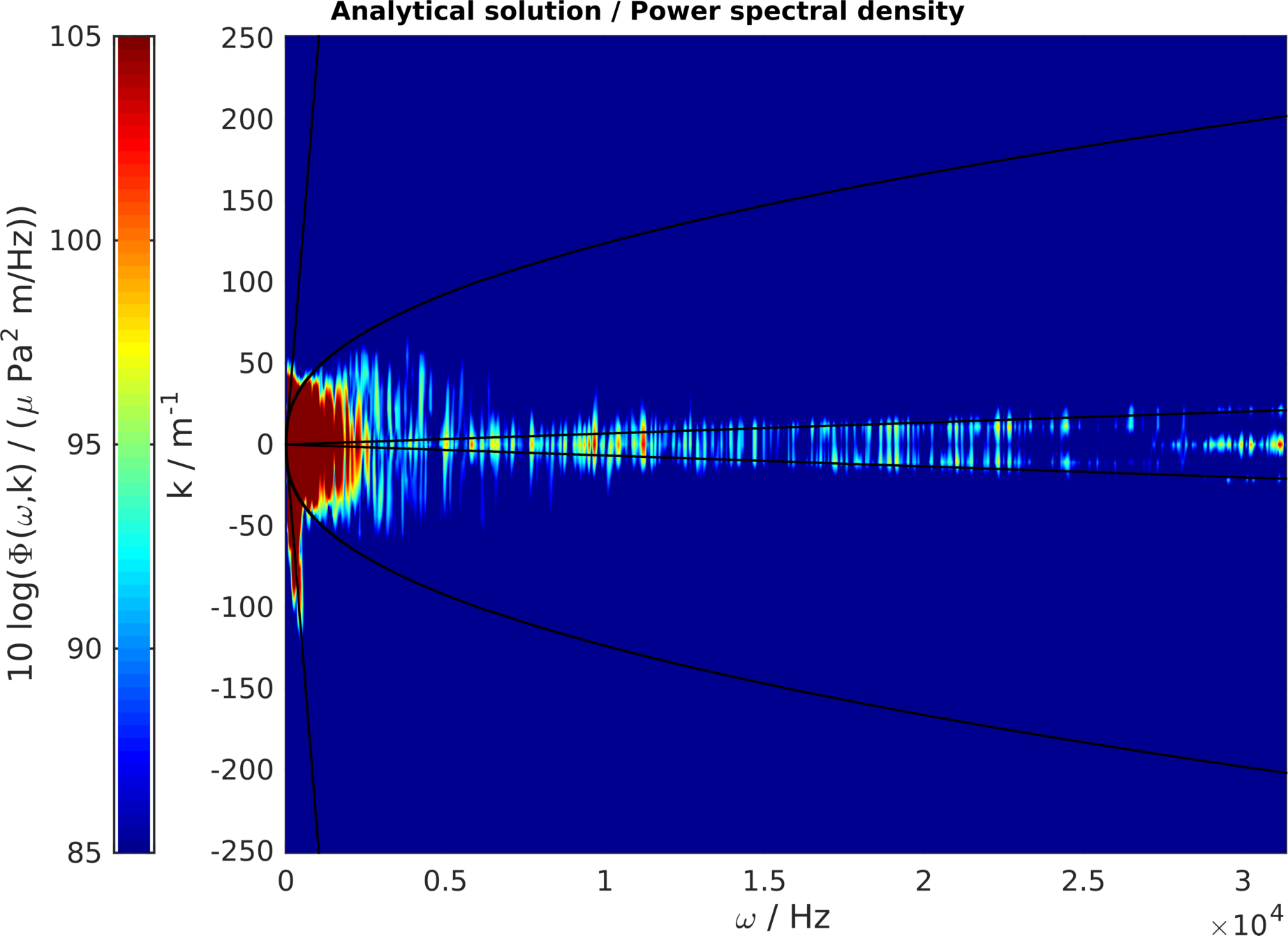}
\hfill
\includegraphics[width=0.475\textwidth]{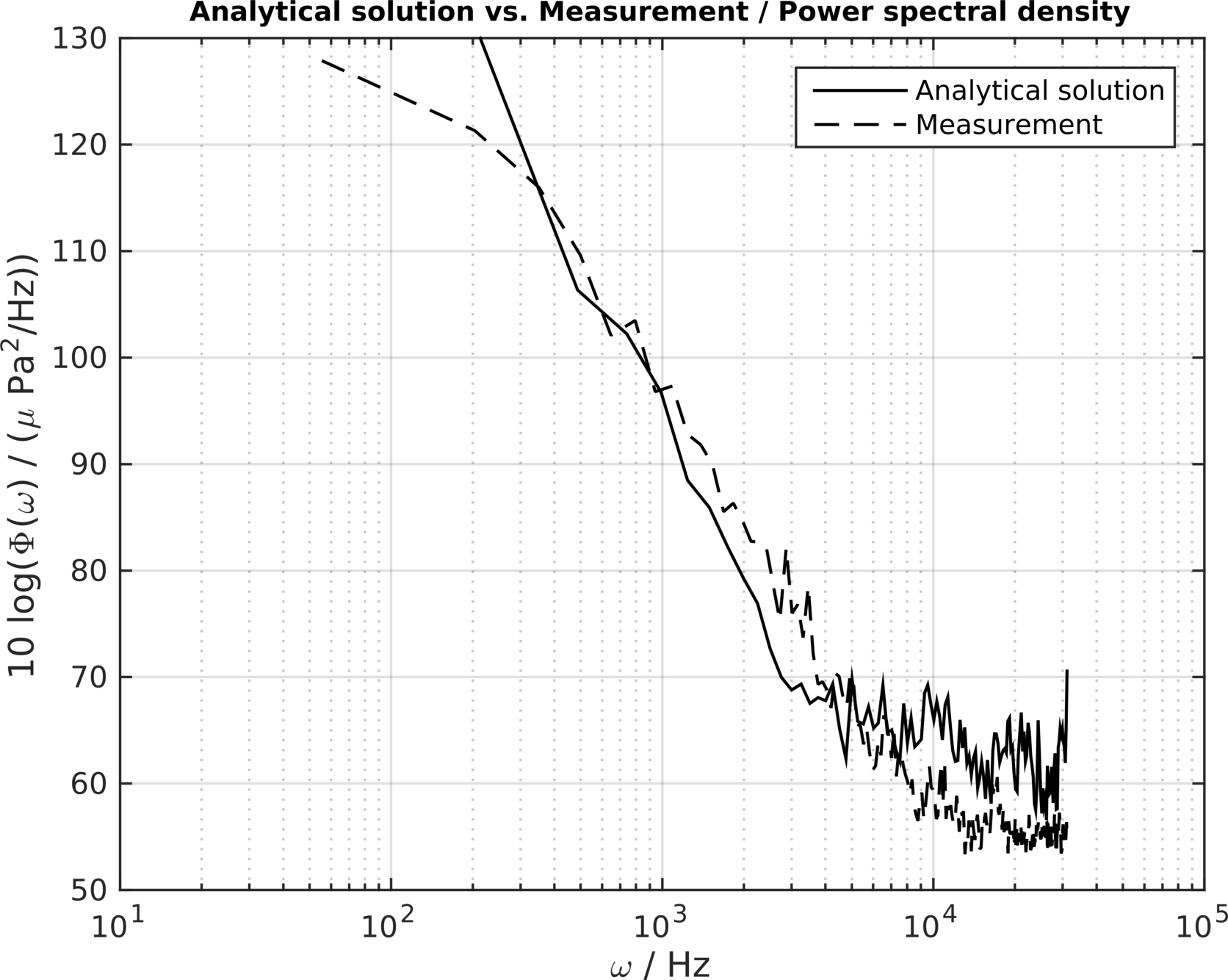}
\end{center}
\caption{Top: Analytical solution without backward waves and with damping $\eta_1=0.05,\,\eta_2=0.25.$ Centre: Isotropic damping $\eta_1=0.05,\,\eta_2=0.05.$ 
Bottom: Isotropic damping $\eta_1=0.25,\,\eta_2=0.25.$}
\label{fig:change_damping}
\end{figure}

\begin{figure}[!htbp]
\begin{center}
\includegraphics[width=0.515\textwidth]{komega_hydrophone_pressure_ref_analytic_damp0_005_025_flattopwin_wob}
\hfill
\includegraphics[width=0.475\textwidth]{energy_spectral_density_ref_analytic_damp0_005_025_flattopwin_wob_sa}
\end{center}

\begin{center}
\includegraphics[width=0.515\textwidth]{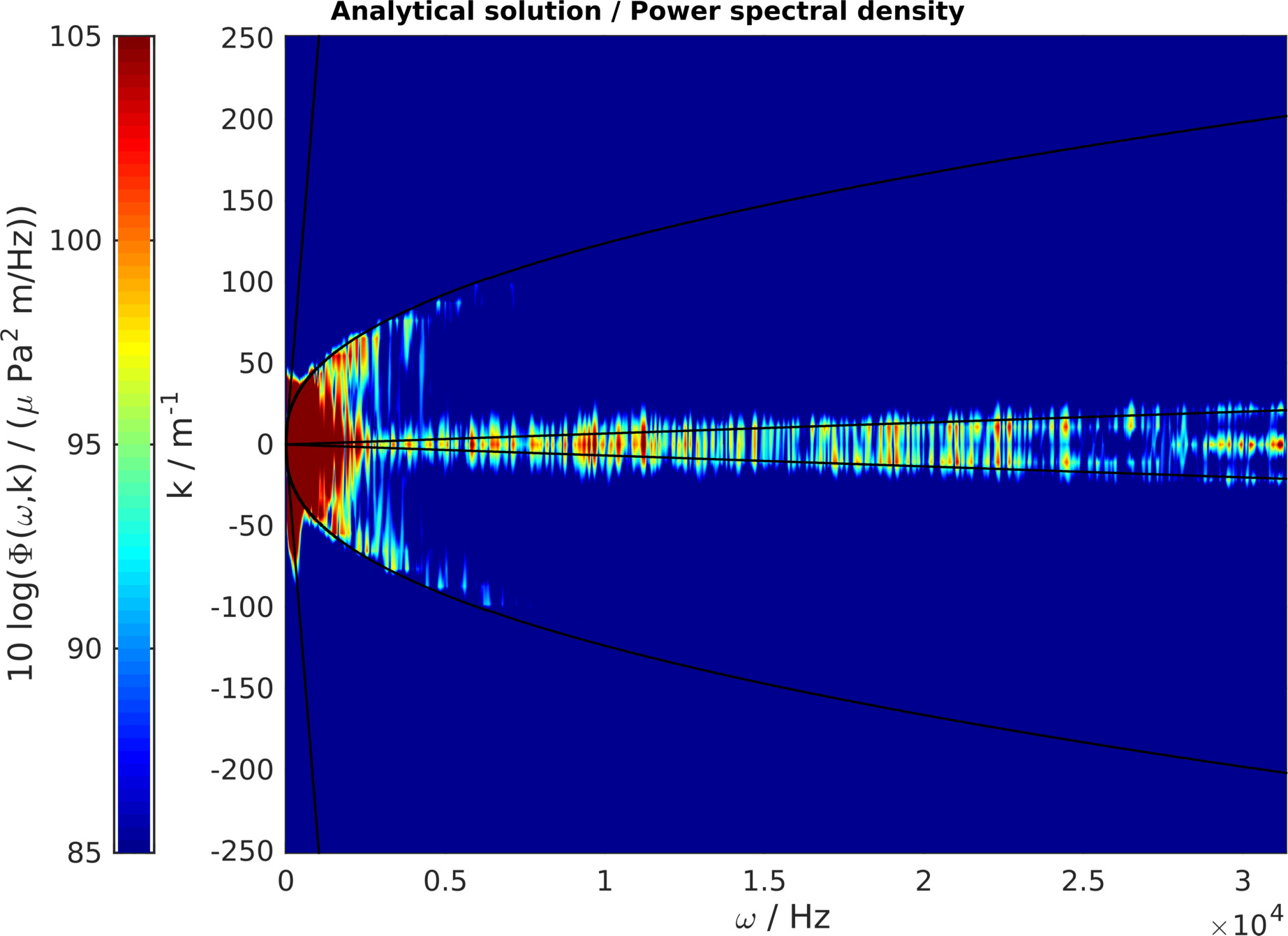}
\hfill
\includegraphics[width=0.475\textwidth]{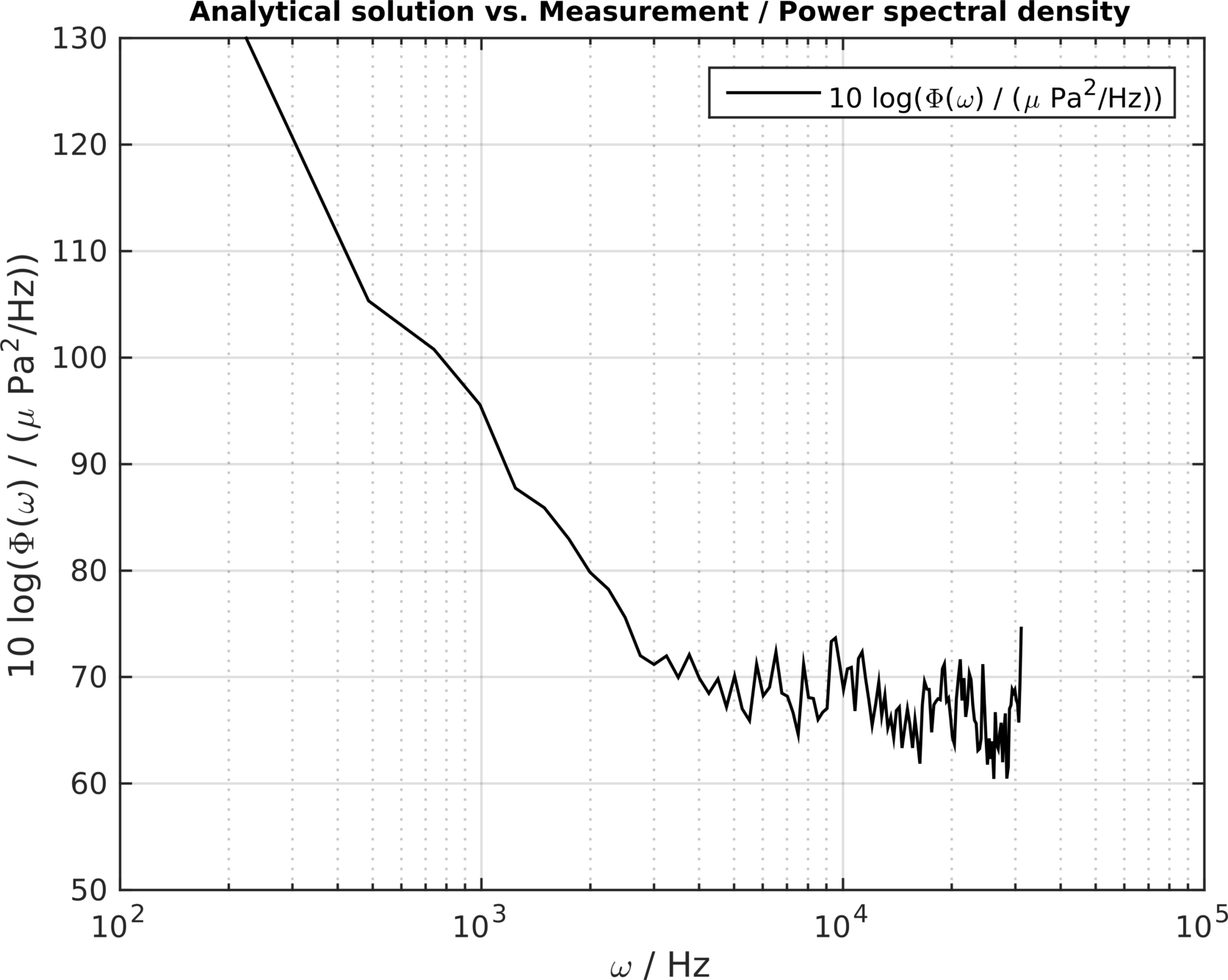}
\end{center}

\begin{center}
\includegraphics[width=0.515\textwidth]{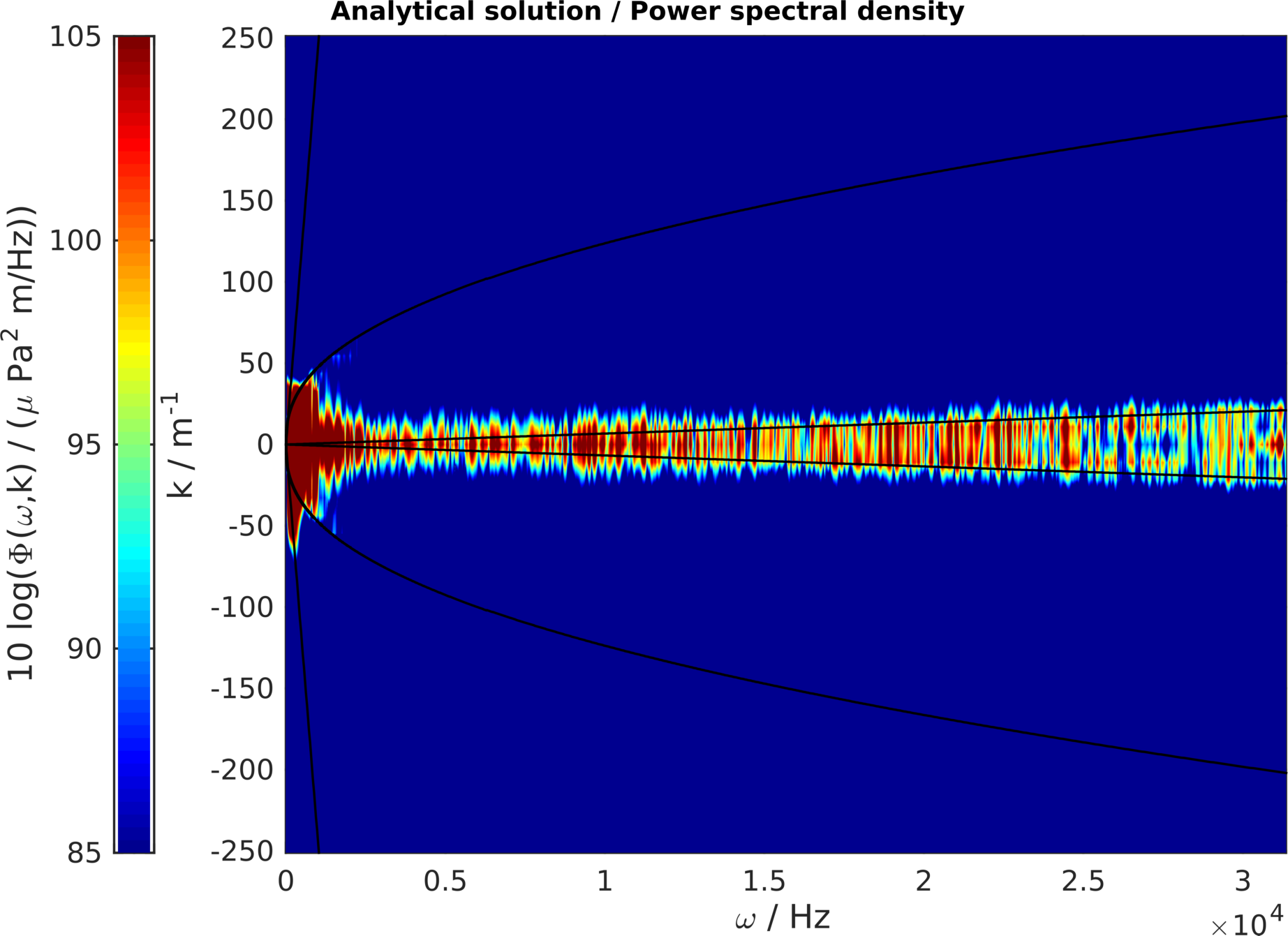}
\hfill
\includegraphics[width=0.475\textwidth]{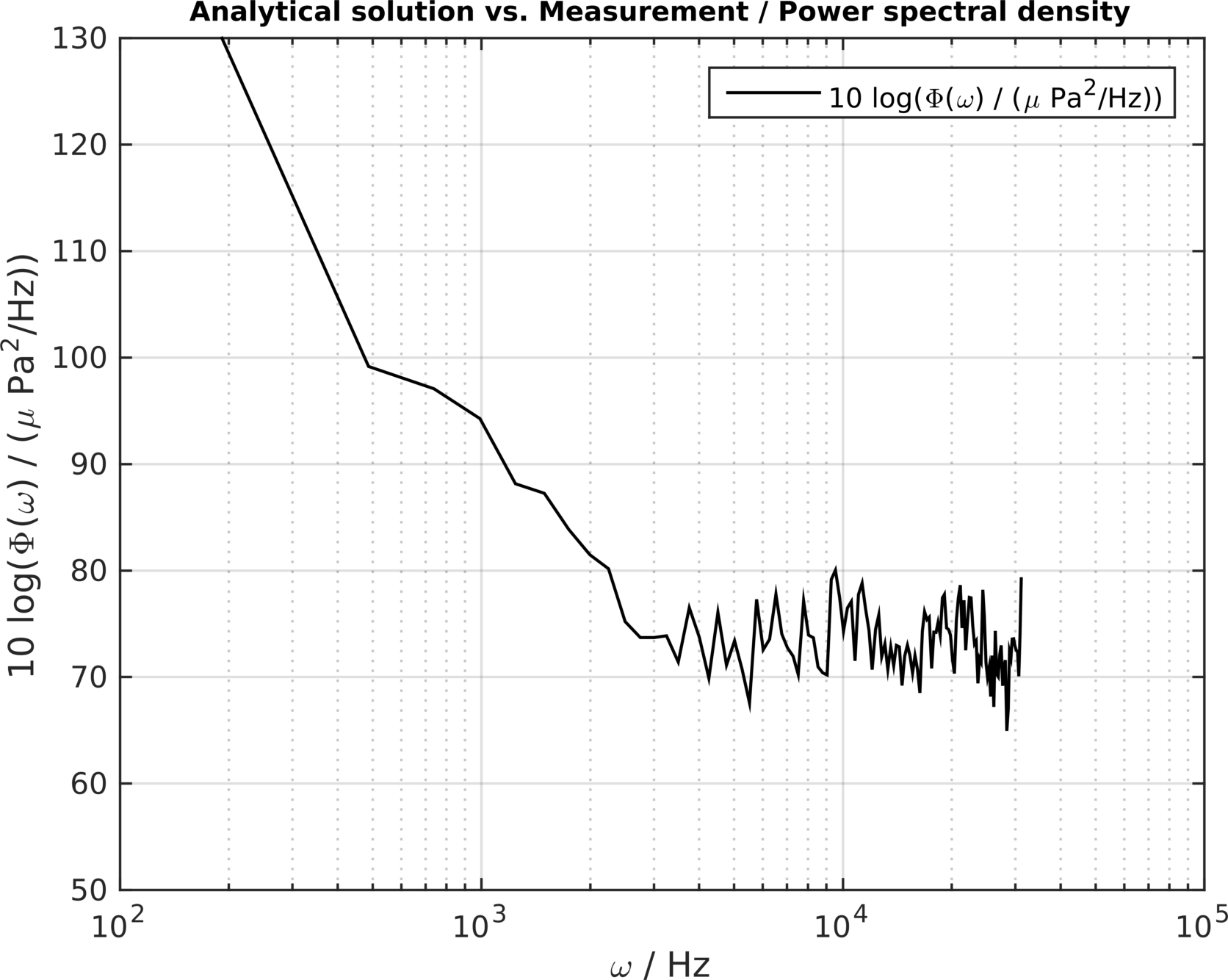}
\end{center}
\caption{Top: Analytical solution without backward waves and with a plate distance $x_3=-0.01$m. Centre: plate distance $x_3=-0.02$m.
\label{fig:change_hydrophone_distance}
Bottom: plate distance $x_3=-0.05$m.}
\end{figure}

\begin{figure}[!htbp]
\begin{center}
\includegraphics[width=0.515\textwidth]{komega_hydrophone_pressure_ref_analytic_damp0_005_025_flattopwin_wob}
\hfill
\includegraphics[width=0.475\textwidth]{energy_spectral_density_ref_analytic_damp0_005_025_flattopwin_wob_sa}
\end{center}
\begin{center}
\includegraphics[width=0.515\textwidth]{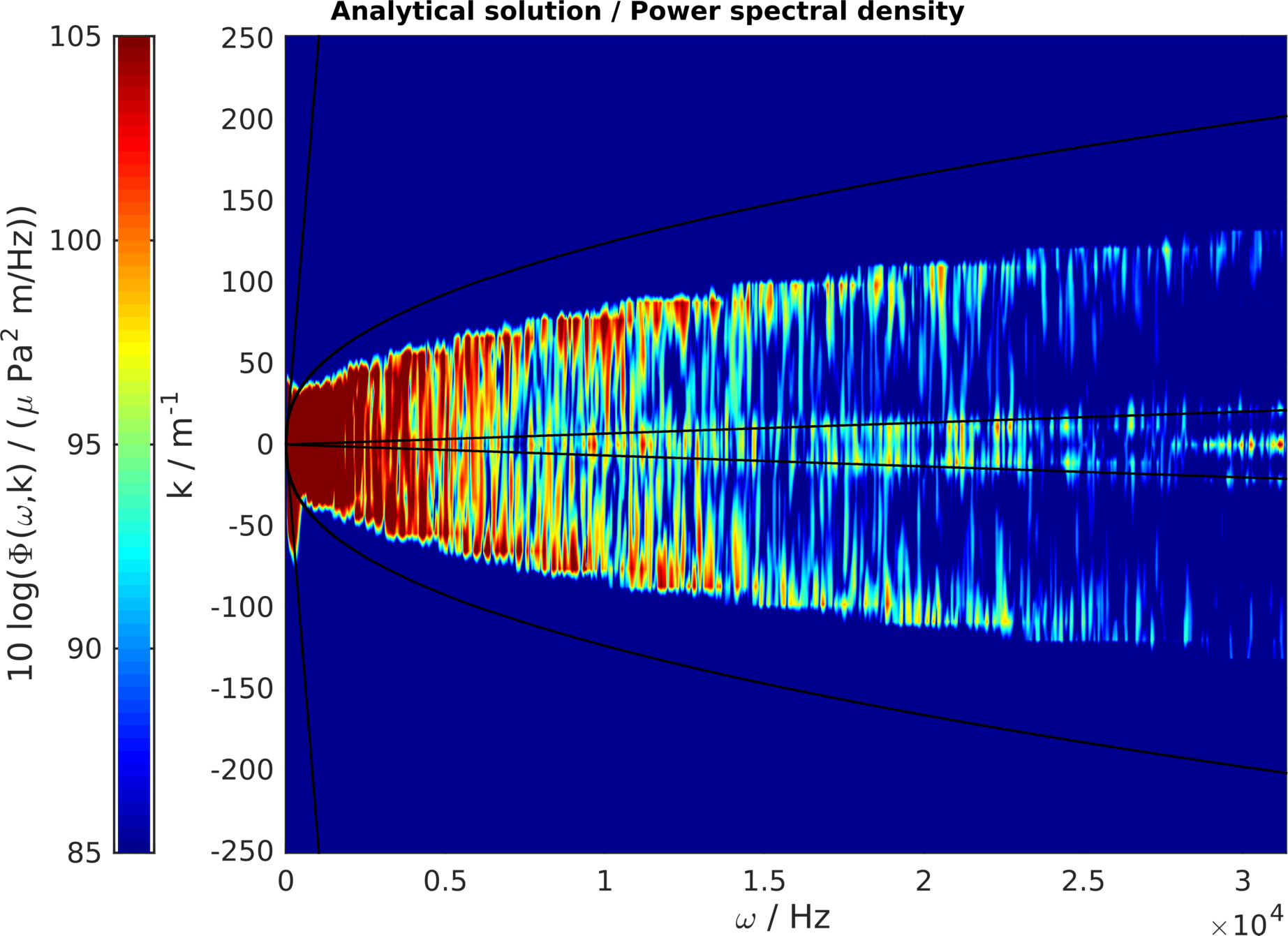}
\hfill
\includegraphics[width=0.475\textwidth]{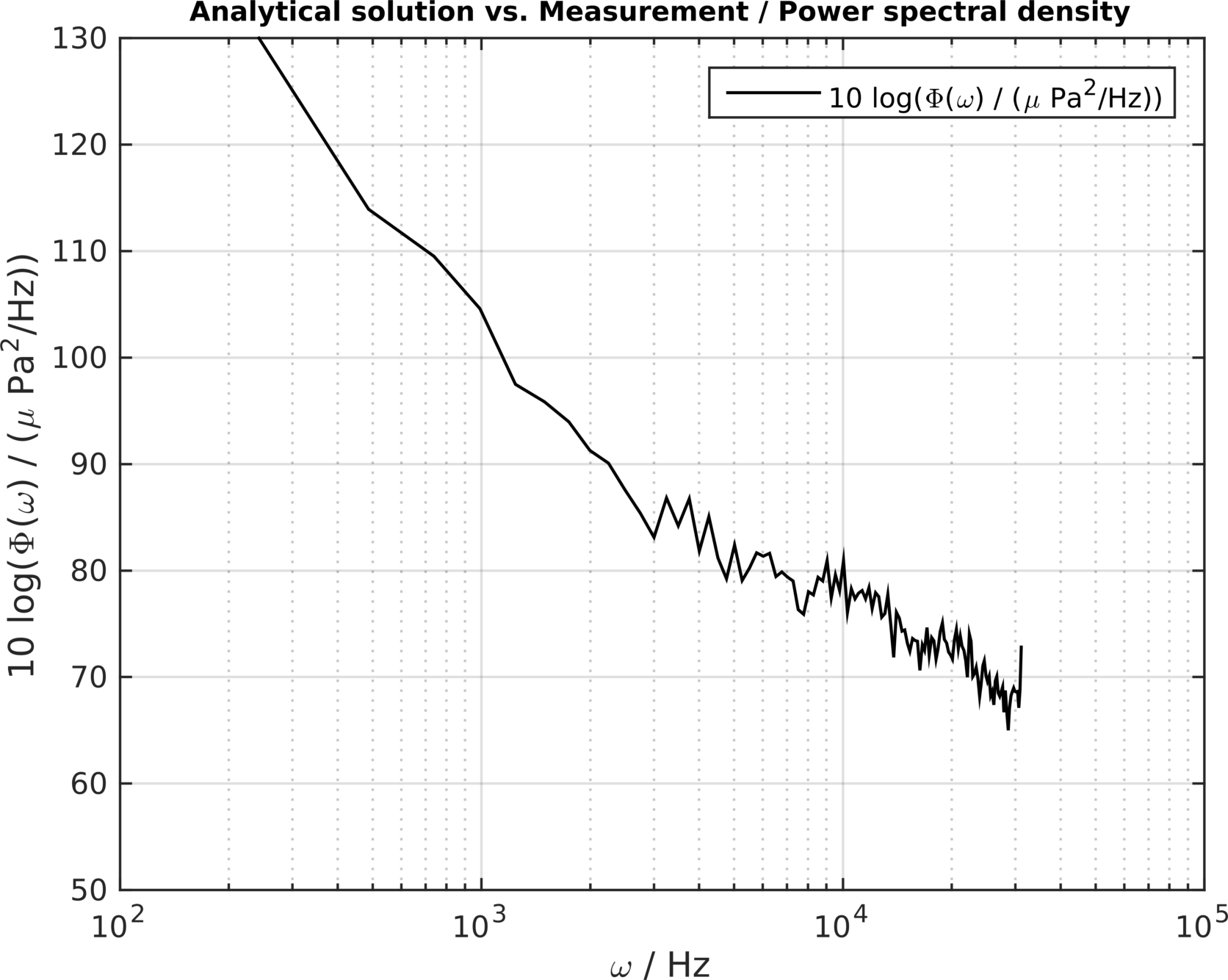}
\end{center}
\begin{center}
\includegraphics[width=0.515\textwidth]{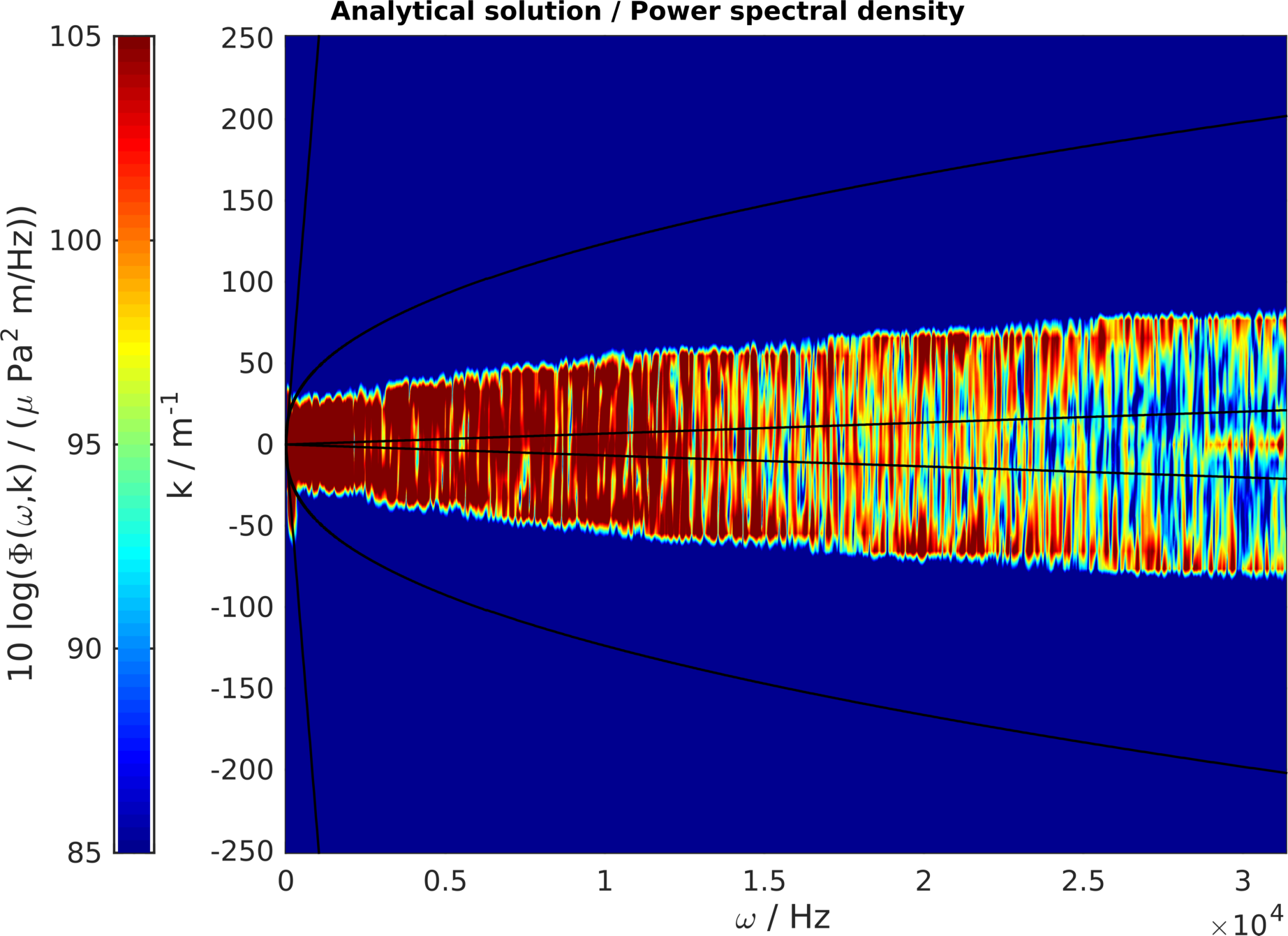}
\hfill
\includegraphics[width=0.475\textwidth]{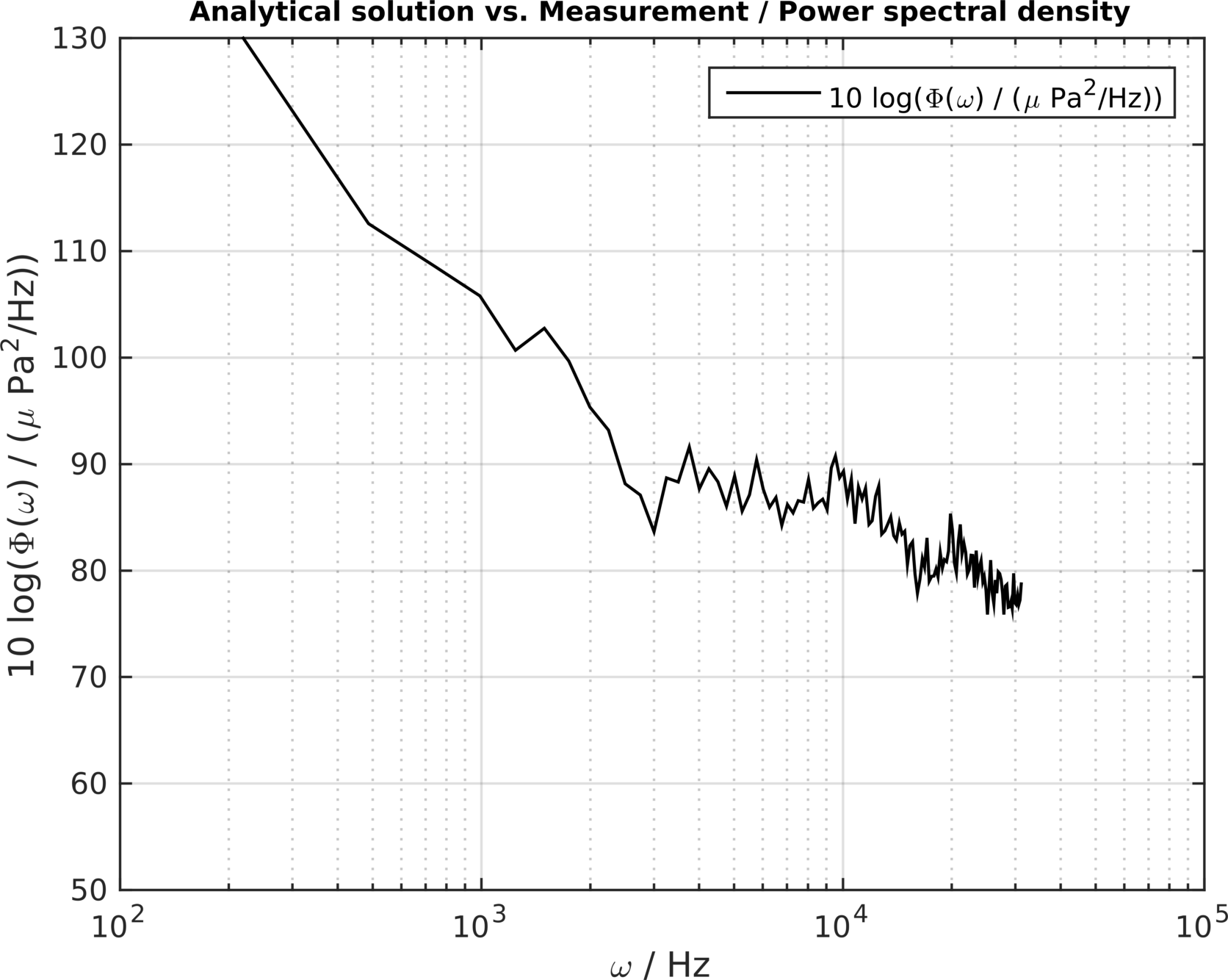}
\end{center}
\caption{Top: Analytical solution without backward waves and with a plate thickness $h=0.0008$m. Centre: plate thickness $h=0.0016$m.
Bottom: plate thickness $h=0.004$m.}
\label{fig:change_plate_thickness}
\end{figure}

The presence of an exact solution allows sensitivity studies of the parameters. In the following the effects of changing the damping parameters, the hydrophone distance and the plate thickness are considered. 
To demonstrate the orthotropic damping effects of the parameters $\eta_1=0.05$ and $\eta_2=0.25$, two isotropic cases $\eta_1=\eta_2=0.05$ and  $\eta_1=\eta_2=0.25$ are shown in Figure \ref{fig:change_damping}. These cases show that the wavenumber-frequency pattern of the sea trial could only be realised by an orthotropic damping.
Moreover, Figure \ref{fig:change_hydrophone_distance} shows the exponentially decay of the bending waves with the distance of the plate. 
Finally, changing the plate thickness is shown in Figure \ref{fig:change_plate_thickness}. Here, the wavenumber-frequency pattern is drawn together to lower wavenumbers and it is found that thicker plates transmits more noise at large frequencies which should be interpreted in the light of the wall pressure frequency resolution (cf. Figure \ref{fig:hydrodynamic_psd}). According to the definition of the plate parameters $m$ and $B,$ the same effect could be realised if Young's modulus will be raised.

\section{Conclusions}
In the present study, it is demonstrated that 
the problem of flow noise generation and propagation in hull-mounted sonar systems could be solved by an numerical/analytical solution approach. The underlying system of partial differential equations considers a hydrodynamical source term and the fluid structure interaction of bending and acoustic waves with respect to different isotropic materials and geometries. These set of equations can also be found in vibro-acoustic applications of thin-walled passenger cabins of cars, trains, ships or aeroplanes. 
In this paper, the analytical solution of the equations under consideration has been derived for the first time without assumptions on the fluid loading. 
By changing the parameters and the source term of the analytical solution the results can be generalised to different rectangular geometries, material parameters, damping ratios, flow speeds. Moreover, if the listener only hears the incoming wave (e.g. shading of reflected waves through a hydrophone holder) then the results can be also applied for non-rectangular geometries.
Finally, the numerical/analytical solutions are compared with the measurement result. 
The perfect agreement in the $\omega^{-4}$-range of the frequency spectrum and the pattern in the wavenumber-frequency plots 
show that the accuracy of the modelling approach is within the scope of sea trial measurement accuracies.

\section*{Acknowledgements}
The author would like to thank Carl Erik Wasberg, Norwegian Defense Research Establishment (FFI), for granting him access to the inflow simulation with obstacles.
Moreover, he would like to thank J.~Abshagen, D.~K\"uter and V.~Nejedl, Bundeswehr Technical Centre for Ships and Naval Weapons (WTD 71), for allowing him to use their unpublished sea trial results.


\newpage

\section*{References}
\begin{thebibliography}{3}
\bibitem[1]{Abshagen.Kueter.ea_Flowinducedinterior_2016}
J.~Abshagen, D.~K\"uter, V.~Nejedl,
\newblock Flow-induced interior noise from a turbulent boundary layer of a towed body.
\newblock Advances in Aircraft and Spacecraft Science, 3(3) (2016) 259--269.
\bibitem[2]{Crighton.Dowling.ea_ModernMethods_1992}
D.G.~Crighton, A.P.~Dowling, J.E.~Ffowcs~Williams,M.~Heckl,F.G.~Leppington,
\newblock Modern Methods in Analytical Acoustics, Lecture Notes.
\newblock Springer-Verlag, 1992.
\bibitem[3]{Howe_AcousticsofFluid-Structure_1998}
M.S.~Howe,
\newblock Acoustics of Fluid-Structure Interactions.
\newblock Cambridge Monographs on Mechanics, 1998.
\bibitem[4]{Ciappi.DeRosa.ea_Flinovia_2015}
E.~Ciappi, S.~De~Rosa, F.~Franco, J.L.~Guyader, S.A.~Hambric,
\newblock Flinovia - Flow Induced Noise and Vibration Issues and Aspects.
\newblock Springer, 2015.
\bibitem[5]{Strawderman_WavevectorFrequencyAnalysis_1994}
W.A.~Strawderman,
\newblock Wavevector-Frequency Analysis with Applications to Acoustics.
\newblock Defense Technical Information Center, 1994.
\bibitem[6]{Graham_HighFrequencyVibration_1995}
W.R.~Graham,
\newblock High Frequency Vibration and Acoustic Radiation of Fluid-loaded Plates.
\newblock Philosophical Transactions of the Royal Society of London, Series A 352 (1995) 1--43.
\bibitem[7]{Graham_BoundaryLayerInduced1_1996}
W.R.~Graham,
\newblock Boundary Layer Induced Noise in Aircraft, Part I: The Flat Plate Model.
\newblock Journal of Sound and Vibration 192(1) (1996) 101--120.
\bibitem[8]{Graham_BoundaryLayerInduced2_1996}
W.R.~Graham,
\newblock Boundary Layer Induced Noise in Aircraft, Part II: The Trimmed Flat Plate Model.
\newblock Journal of Sound and Vibration 192(1) (1996) 121--138.
\bibitem[9]{Efimtsov_Characteristics_1982}
B.M.~Efimtsov,
\newblock Characteristics of the field of turbulent wall pressure fluctuations at large Reynolds numbers.
\newblock Soviet Physics Acoustics 28(4) (1982) 289--292.
\bibitem[10]{Maury.Gardonio.ea_WavenumberApproachModelling1_2002}
C.~Maury, P.~Gardonio, S.J.~Elliott,
\newblock A Wavenumber Approach to Modelling the Response of a Randomly Excited Panel, Part I: General Theory.
\newblock Journal of Sound and Vibration 252(1) (2002) 83--113.
\bibitem[11]{Maury.Gardonio.ea_WavenumberApproachModelling2_2002}
C.~Maury, P.~Gardonio, S.J.~Elliott,
\newblock A Wavenumber Approach to Modelling the Response of a Randomly Excited Panel, Part II: Application to Aircraft Panels excited by a Turbulent Boundary Layer. 
\newblock Journal of Sound and Vibration 252(1) (2002) 115--139.
\bibitem[12]{Mazzoni.Kristiansen_FiniteDifferenceMethodAcouticRadiation_1998}
D.~Mazzoni, U.~Kristiansen,
\newblock Finite Difference Method for the Acoustic Radiation of an Elastic Plate Excited by a Turbulent Boundary Layer: a Spectral Domain Solution.
\newblock Flow, Turbulence and Combustion 61 (1998) 133--159.
\bibitem[13]{Mazzoni_EfficientApproximationVibroAcoutic_2003}
D.~Mazzoni,
\newblock An Efficient Approximation for the Vibro-Acoustic Response of a Turbulent Boundary Layer Excited Panel.
\newblock Journal of Sound and Vibration 264(4) (2003) 951--971.
\bibitem[14]{Lebedev_IntroductionFunctionalAnalysis_1997}
V.I.~Lebedev,
\newblock An Introduction to Functional Analysis in Computational Mathematics.
\newblock Birkhäuser Boston, 1997.
\bibitem[15]{Blake_MechanicsofFlow-Induced_1986}
W.K.~Blake,
\newblock Mechanics of Flow-Induced Sound and Vibration, Volume II.
\newblock Academics Press, INC., 1986.
\bibitem[16]{Wagner.Huettl.ea_Large-EddySimulationAcoustics_2007}
C.~Wagner, T.~Hüttl, P.~Sagaut,
\newblock Large-Eddy Simulation for Acoustics.
\newblock Cambridge University Press, 2007.
\bibitem[17]{Hambric.Hwang.ea_VibrationsPlates_2004}
S.A.~Hambric, Y.F.~Hwang, W.K.~Bonness
\newblock Vibrations of plates with clamped and free edges excited by low-speed turbulent boundary layer flow.
\newblock Journal of Fluids and Structures 19 (2004) 93--110.
\bibitem[18]{Xing.Liu_NewExactSolutionsFreeVibrations_2009}
Y.F.~Xing, B.~Liu,
\newblock New exact solutions for free vibrations of thin orthotropic rectangular plates.
\newblock Composite Structures 89 (2009) 567--574.
\end{thebibliography}
\end{document}